\documentclass[11pt,a4paper,usenames]{article}
\usepackage{jheppub}
\usepackage{dsfont}
\usepackage{amsbsy}
\usepackage{bigstrut}
\usepackage{ulem}

\def\be{\begin{equation}}
\def\ee{\end{equation}}
\def\ba{\begin{eqnarray}}
\def\ea{\end{eqnarray}}
\newcommand{\non}{\nonumber \\}
\newcommand{\alsub}{{\boldsymbol{\alpha}}}
\newcommand{\alpsub}{{\boldsymbol{\alpha'}}}

\let\oldhat\hat
\renewcommand{\vec}[1]{\mathbf{#1}}
\renewcommand{\hat}[1]{\oldhat{#1}}

\title{Finite volume effects in pion-kaon scattering\\ and reconstruction of the
$\boldsymbol{\kappa}$(800) resonance}
\author[a]{M. D\"oring}
\author[a,b]{and U.-G. Mei{\ss}ner}

\affiliation[a]
{Universit\"{a}t Bonn, Helmholtz-Institut f\"{u}r
Strahlen- und Kernphysik (Theorie) and Bethe Center for Theoretical 
Physics, D-53115 Bonn, Germany}

\affiliation[b]
{Forschungszentrum J\"{u}lich, Institut f\" {u}r Kernphysik, Institute for 
Advanced Simulation and J\"ulich Center for Hadron Physics,
D-52425 J\"{u}lich, Germany}

\emailAdd{doering@hiskp.uni-bonn.de}
\emailAdd{meissner@hiskp.uni-bonn.de}

\abstract{
Simulating the $\kappa(800)$ on the lattice is a challenging task that starts to become
feasible due to the rapid progress in recent-years lattice QCD calculations. As the
resonance is broad, special attention to finite-volume effects has to be paid, because no
sharp resonance signal as from avoided level crossing can be expected. In the present
article, we investigate the finite volume effects in the framework of unitarized chiral
perturbation theory using next-to-leading order terms. After a fit to meson-meson partial
wave data, lattice levels for $\pi K$ scattering are predicted. In addition, levels are
shown for the quantum numbers in which the $\sigma(600)$, $f_0(980)$, $a_0(980)$,
$\phi(1020)$, $K^*(892)$, and $\rho(770)$ appear, as well as the repulsive channels. Methods
to extract the $\kappa(800)$ signal from the lattice spectrum are presented. Using
pseudo-data, we estimate the precision that lattice data should have to allow for a
clear-cut extraction of this resonance. To put the results into context, in particular the
required high precision on the lattice data, the $\sigma(600)$, the $P$-wave resonances
$K^*(892)$ and $\rho(770)$, and the repulsive $\pi K,\,\pi\pi$ phases are analyzed as well.
}

\notoc
\begin{document}
\maketitle


\section{Introduction}
Scalar resonances  attract increasing interest as, with the rapid progress of lattice QCD
simulations~\cite{Durr:2008zz}, the excited resonance spectrum starts to come into
reach~\cite{Blossier:2007vv,Lin:2008pr,Gockeler:2008kc,Gattringer:2008vj,Dudek:2010wm,Baron:2010bv,Engel:2010my,Feng:2010es,Dudek:2010ew,Dudek:2011tt,Morningstar:2011ka,Beane:2011sc,Morningstar:2011nt,Lang:2011mn,Menadue:2011pd,Fu:2011xw},
including first results on the $\kappa(800)$~\cite{Fu:2011xw}. The strangeness $S=-1$,
$\kappa(800)$ resonance is of particular interest because it does not mix with scalar
glueballs and is thus expected to reveal itself with a clearer signal than, e.g., the broad
$\sigma(600)$ in the scalar-isoscalar sector. On the other hand, the $\kappa$ presents a
similar challenge, both for experimental and lattice studies, as this resonance is similarly
broad. For wide resonances, the {\it avoided level crossing} of the lattice levels is washed
out\footnote{This phenomenon was first studied in Ref.~\cite{Bernard:2007cm}
  in the context of extracting the $\Delta$-resonance properties.} 
and a clear-cut extraction of the resonance pole and its properties requires additional
effort. Such complications from finite volume effects can be addressed in the framework of
field theory with periodic boundary conditions, as pioneered by
L\"uscher~\cite{Luscher:1986pf,Luscher:1990ux,Luscher:1991cf}. On one hand, this framework
allows to predict lattice levels using hadronic approaches originally set up for the
infinite volume limit. On the other hand, phase shifts, and in consequence resonance
properties, can be extracted from existing lattice data. In the absence of such data, as it
is the case for the $\kappa(800)$, the formalism can still be used to study the propagation
of the unavoidable uncertainties from lattice data to the extracted resonance properties.
Predicting lattice levels in the meson-meson sector, using the chiral expansion up to
next-to-leading order (NLO) terms, and determining the needed precision of the  lattice
spectrum to make sensible statements about the $\kappa(800)$, is the purpose of this study.

For the prediction of lattice levels and the generation of pseudo-data, the chiral unitary
scheme of ref.~\cite{Oller:1998hw} is employed that includes terms up to NLO in the chiral
expansion. Chiral Perturbation Theory (CHPT)~\cite{Gasser:1983yg,Gasser:1984gg} has been
applied to the problem of $\pi K$ scattering since a long time, calculating threshold
parameters~\cite{Bernard:1990kx}, one-loop corrections~\cite{Bernard:1990kw}, and extensions
to higher energies by including resonances~\cite{Bernard:1991zc}.  The unitarization of the
amplitude can induce strong non-linearities leading to non-perturbative features such as pole
formation and has been explored, e.g., in ref.~\cite{Oller:1997ti} where the $\sigma(600)$,
$f_0(980)$, and $a_0(980)$ resonances appear dynamically generated from the unitarization of
the lowest order (LO) chiral meson-meson interaction in a Bethe-Salpeter type of equation.
This approach has been further developed and refined by, e.g., taking resonances into account
to extend the applicability to higher
energies~\cite{Oller:1998zr,Jamin:2000wn,Albaladejo:2008qa,Guo:2011pa}. An alternative scheme
(among other approaches such as meson exchange models~\cite{Janssen:1994wn,Krehl:1996rk}) is
provided by the inverse amplitude method  ---the approach employed
here~\cite{Oller:1998hw,Dobado:1992ha,Oller:1997ng,Nieves:2001de}. It was extended to provide
the full one-loop calculation for the chiral transition potential in
SU(3)~\cite{GomezNicola:2001as} and to study the quark-mass dependence of
resonances~\cite{Hanhart:2008mx,Nebreda:2010wv,Nebreda:2011xw}. Dispersive analyses of
meson-meson scattering as formulated, e.g., in Roy equations provide a related
model-independent approach, often stabilized by the use of chiral perturbation
theory~\cite{Ananthanarayan:2000ht,Caprini:2003ta,Buettiker:2003pp,DescotesGenon:2006uk,DescotesGenon:2007ta,Hoferichter:2011vq}. 

To learn about the $\kappa(800)$ resonance on the lattice, the mentioned approaches, all of
them set up for the infinite volume limit, can be extended to predict lattice levels using
L\"uscher's formalism. The idea of using effective field theory to study the lattice
spectrum and excited states has been been developed for the $\Delta(1232)$ resonance and
other meson-baryon
systems~\cite{Bernard:2007cm,Bernard:2008ax,Bernard:2009mw,Lage:2009zv,Hoja:2010fm} as well
as for the meson sector~\cite{Bernard:2010ex,Bernard:2010fp,Doring:2011vk}. A special role
plays here the generalization of the L\"uscher formulation to coupled
channels~\cite{Lage:2009zv,Bernard:2010fp,Doring:2011vk}: e.g., with the quantum numbers of
the $\kappa(800)$, one has the $\pi K$ and the $\eta K$ channels, the latter not being too far
away from the $\kappa(800)$ pole, thus potentially influencing its properties. In such
situations, phases and resonances can still be extracted using a $K$-matrix
approach~\cite{Bernard:2010fp} together with the L\"uscher ${\cal Z}_{00}$ function or
directly the relativistic meson-meson propagator~\cite{Doring:2011vk}. Compared to the
one-channel case, for which the L\"uscher formula is
valid~\cite{Luscher:1986pf,Luscher:1990ux}, the phase extraction in the two-channel case
requires in general much more precise lattice data as recently shown in
ref.~\cite{Doring:2011vk}, in particular if a particle threshold coincides with a resonance,
as, e.g., the $f_0(980)$ (see also ref.~\cite{Torres:2011pr} for an example in the charm
sector).  Concerning the extraction of higher partial waves (see, e.g.,
refs.~\cite{Bernard:2008ax,Luu:2011ep}) in the presence of multiple channels, another
possible analysis tool for lattice data is given by dynamical coupled-channel
approaches~\cite{Doring:2010ap} that were recently shown to be easily adaptable to the
finite volume problem~\cite{Doring:2011ip}.

In the present study, lattice levels will be constructed and analyzed following the
discretization procedure of ref.~\cite{Doring:2011vk}. The relativistic propagator $G$ used in
that unitary chiral (UCHPT) approach leads to a discretized propagator $\tilde G$ that
provides a formalism equivalent to the L\"uscher approach, plus relativistic effects of order
$e^{-M_\pi\,L}$, the size of which have been discussed in ref.~\cite{Doring:2011vk}. Apart
from these particular $e^{-M_\pi\,L}$ effects, there are of course, effects of the same order
from, e.g., omitted loops in the $t$-channel or polarization effects of the
pion~\cite{Luscher:1991cf}. Note that the propagator $G$ used in ref.~\cite{Oller:1998hw} is
just the same as that of ref.~\cite{Doring:2011vk}. The present work provides an extension of
methods developed in ref.~\cite{Doring:2011vk} to the $\pi K$ sector and goes beyond it in
several aspects: first, the use of NLO terms in the chiral expansion implies higher powers of
the scattering energy $E$ in the potential. In fits to pseudo-data generated from this
interaction, the non-zero higher order terms will present a new challenge for the resonance
extraction and new tools are required. The key is here that the well-known LO chiral
interaction can be used as explicit input in the expansion of the general fit potential, and
the unknown higher order effects can be cast in terms of a systematic expansion in powers of
$s\equiv E^2$. As will be shown, the chiral input helps to stabilize the extraction process
considerably. Also, we formulate a criterion of when to cut this expansion in $s$ in the
resonance extraction process, guided by statistical arguments.

It should be stressed that the proposed reconstruction method of the infinite volume limit
is model-independent in the sense that it does not make assumptions on the existence and 
form of the resonance propagator as, e.g., in
refs.~\cite{Gockeler:2008kc,Feng:2010es,Lang:2011mn,Fu:2011xw}, sometimes in form of an
effective range formula~\cite{Gockeler:2008kc,Feng:2010es}. In particular, the low-lying
scalar resonances are known to be very
broad~\cite{Nakamura:2010zzi,Black:1998zc,Aitala:2002kr,Ablikim:2004qna,Zheng:2003rw,Bugg:2005xx}
and can by no means be approximated by Breit-Wigner or related functions. 

In section~\ref{sec:inifinitev} we briefly introduce the inverse amplitude method of
ref.~\cite{Oller:1998hw} and show the result of our fit  to the meson-meson partial waves
with special weight on the ($\eta K,\,\pi K$) system in which the $\kappa(800)$ and the
$K^*(892)$ reside. The lattice levels, predicted from this solution, are shown in
section~\ref{sec:predictions}. Having set up an amplitude that describes well the $\pi K$
scattering data, with low-energy constants well constrained from various different channels,
the main issue of the paper can be addressed in section~\ref{sec:results}: depending on the
precision of the lattice data, how well can the properties of the $\kappa(800)$ resonance be
determined?  After a discussion of a suitable expansion potential in
section~\ref{sec:whichpotential} ---see the above remarks--- this question is addressed in
section~\ref{sec:higher_pole}, with pseudo-data derived from the global fit obtained before.
As the pole position in this fit is rather at the higher end of the accepted pole
positions~\cite{Nakamura:2010zzi}, we will analyze in section~\ref{sec:lowkappa} the
opposite case with a rather light $\kappa(800)$ as obtained, e.g., in the data analysis
using Roy-Steiner equation of ref.~\cite{DescotesGenon:2006uk}.  As the required precision
of the lattice data, to determine the properties of the $\kappa(800)$, is quite high, one
may also wonder to which extent one can at least make qualitative statements about its
properties, using only one data point from the lowest level. This question is addressed in
section~\ref{sec:lowest}.  To put the results of the $\kappa(800)$ into context and compare
to the required precision for other quantum numbers, section~\ref{sec:other} is dedicated to
the analysis of the $\sigma(600)$ resonance. In particular, we check whether the required
precision as found in ref.~\cite{Doring:2011vk} is modified in the present formalism in
which NLO terms play an important role. In the same section, we compare the results also to
an analysis of the $K^*(892)$ and $\rho(770)$ resonances as well as to the repulsive $\pi K$
and $\pi\pi$ phase shifts. The residues of all analyzed resonances are collected in
section~\ref{sec:residues}. We briefly summarize our conclusions in section~\ref{sec:con}.


\section{The pion-kaon amplitude in the infinite volume limit\\ with next-to-leading order
terms}
\label{sec:inifinitev}

Organizing the LO and NLO contact terms in the inverse amplitude method as derived in
ref.~\cite{Oller:1998hw}, and projecting the interaction to $S$- and $P$-waves, a large body
of partial wave data can be described after a fit of the low-energy constants $L_1$ to
$L_6+L_8$. This successful approach has been refined over the years, see, e.g.,
ref.~\cite{GomezNicola:2001as} including the full one-loop calculation and later also two loop
corrections~\cite{Nebreda:2011xw}. In this study we use the earlier treatment of
ref.~\cite{Oller:1998hw} that is sufficient for the present purposes. Note also that the
loops in the $t$-channel will only provide contributions to the potential that are
exponentially suppressed ($e^{-M_\pi\,L}$ effects) and thus expected to be of minor
importance in the study of finite volume effects (for sufficiently large $M_\pi L$). 

\begin{figure}
\begin{center}
\includegraphics[width=0.49\textwidth]{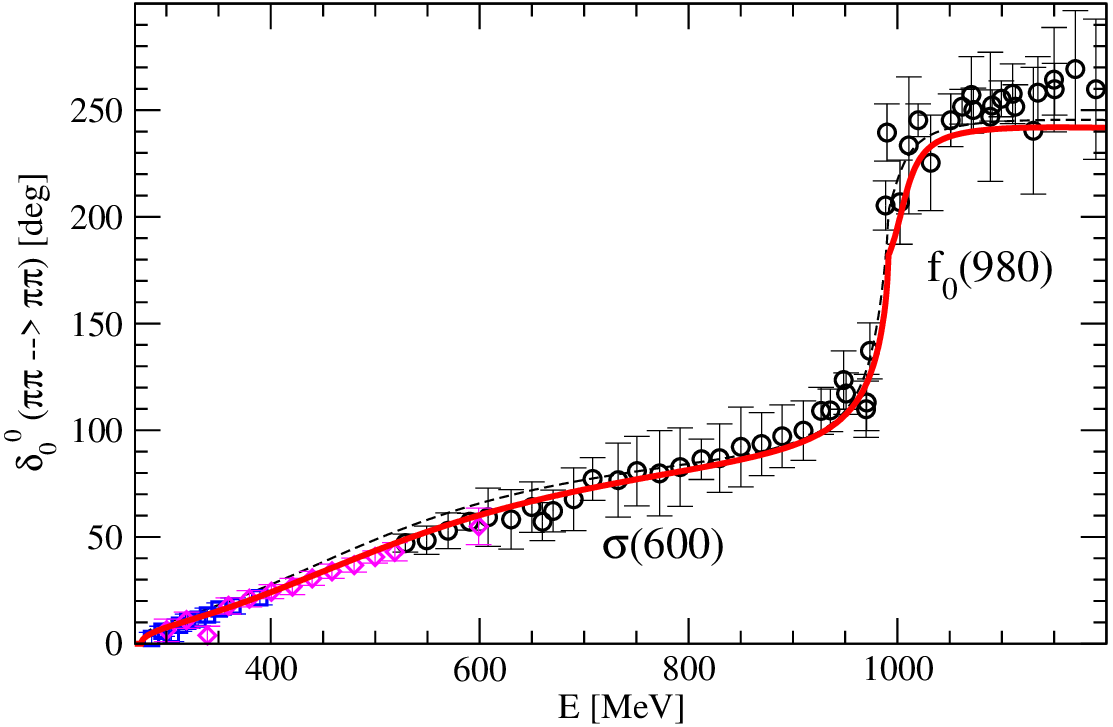}
\includegraphics[width=0.49\textwidth]{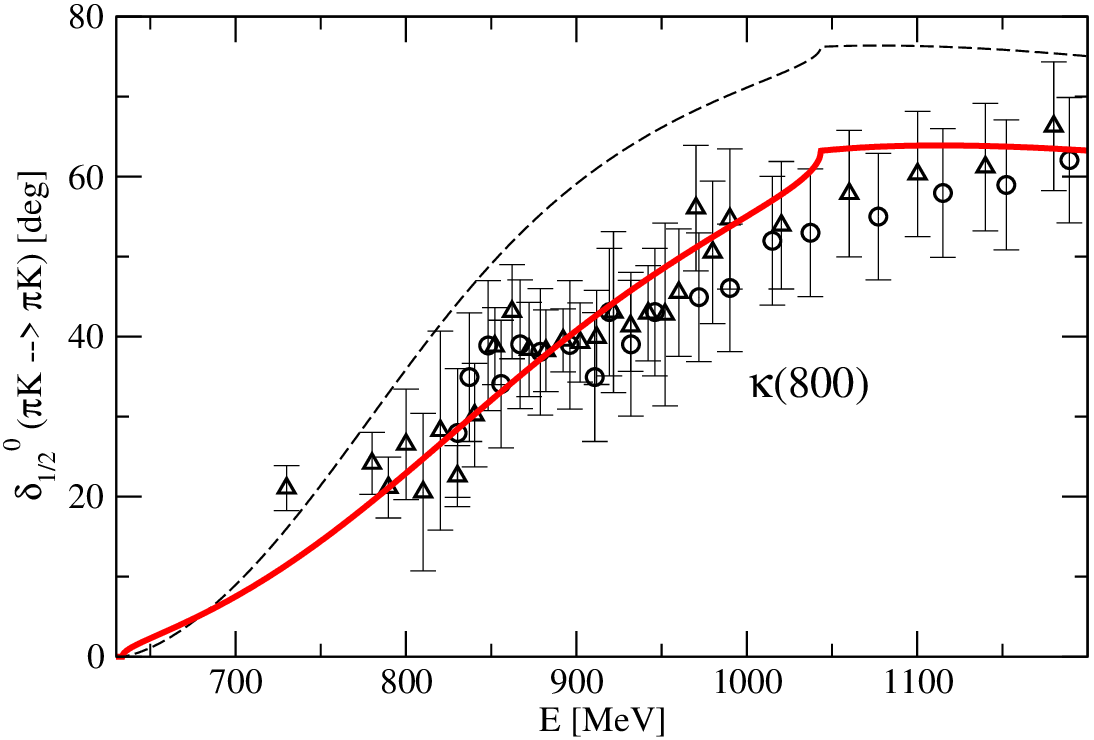} \\
\includegraphics[width=0.49\textwidth]{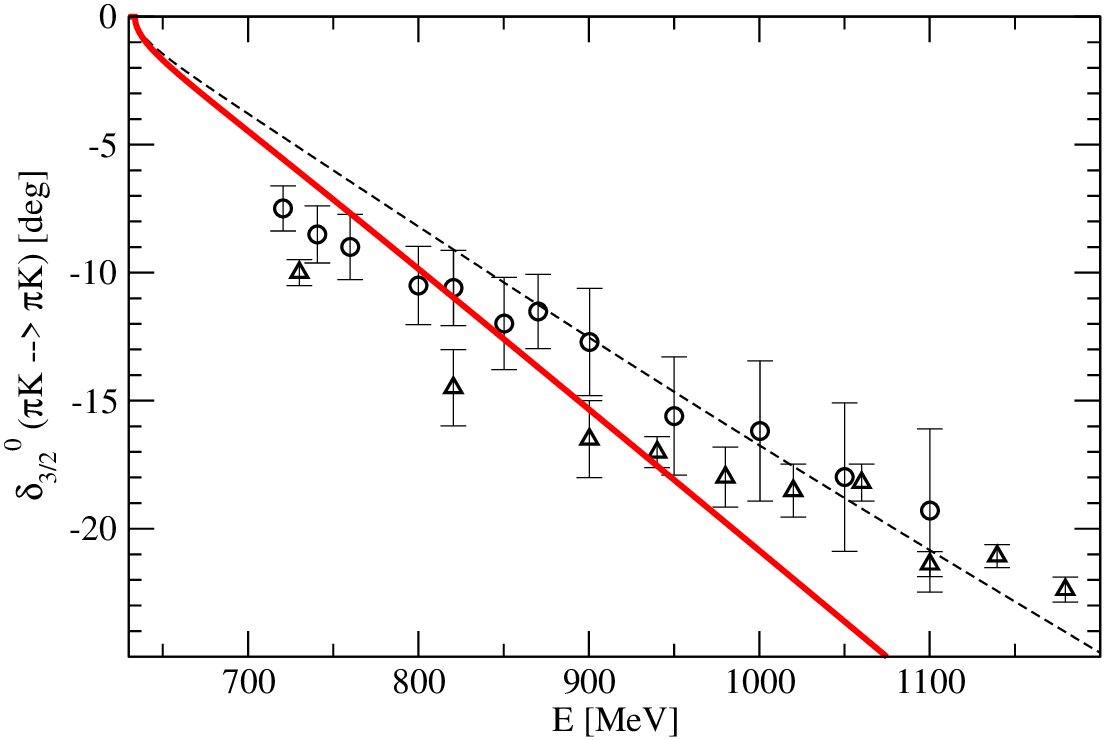} 
\includegraphics[width=0.49\textwidth]{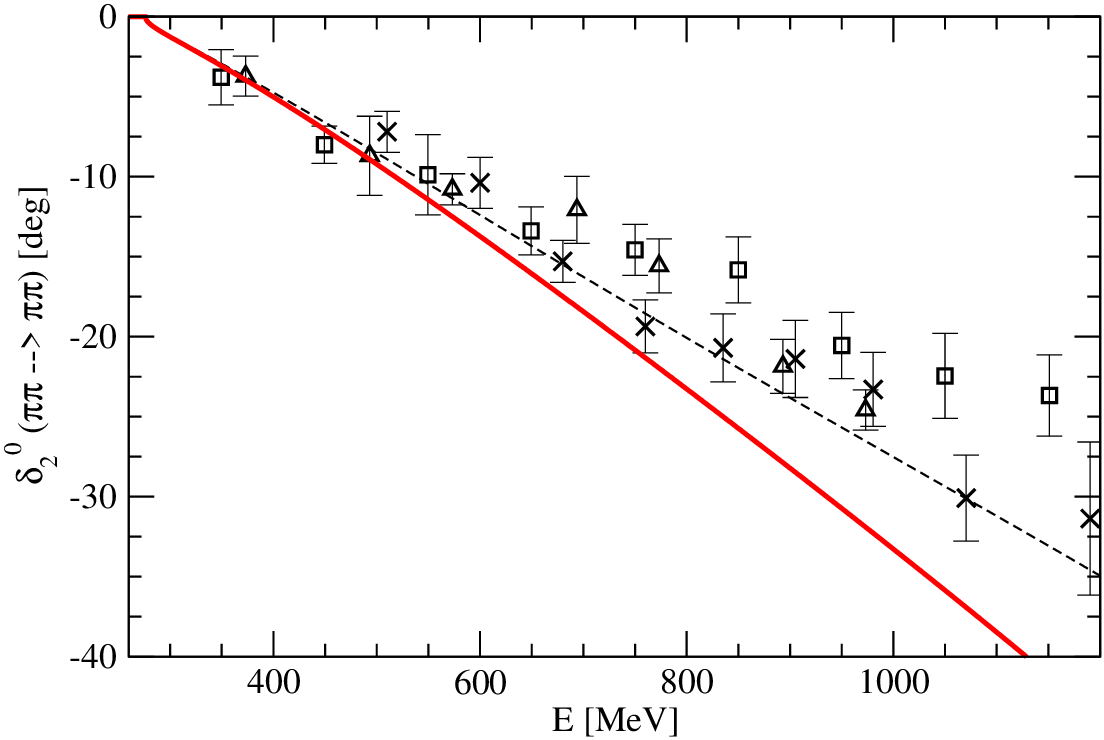} \\
\includegraphics[width=0.49\textwidth]{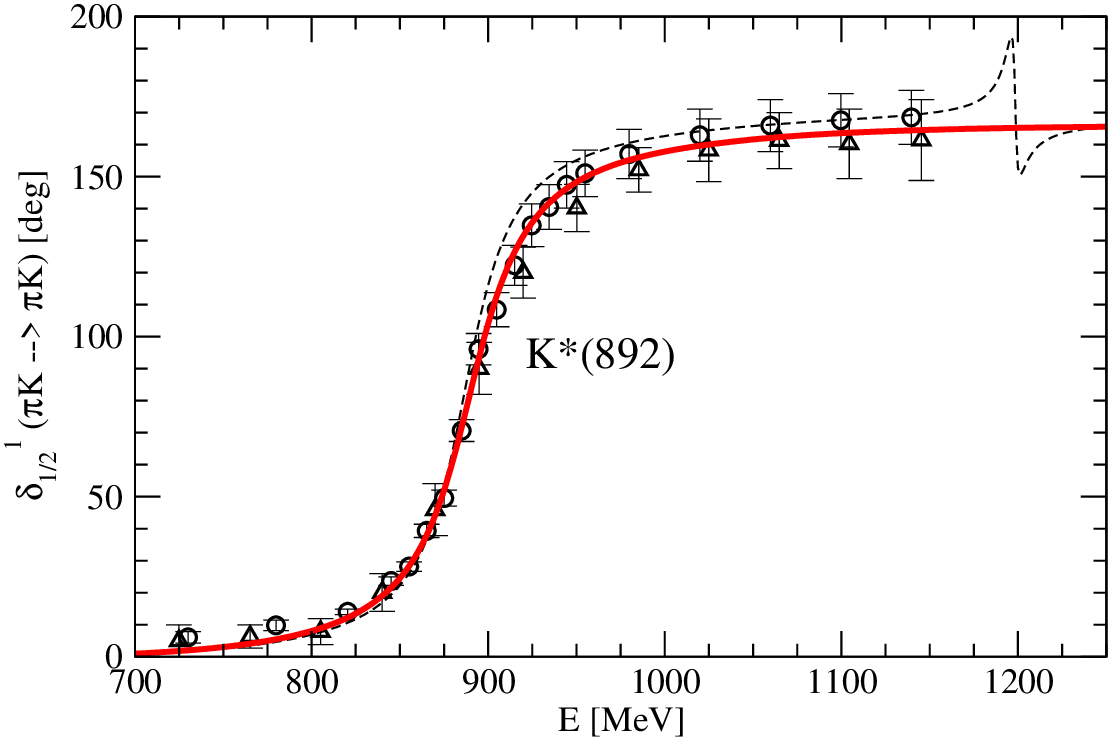} 
\includegraphics[width=0.49\textwidth]{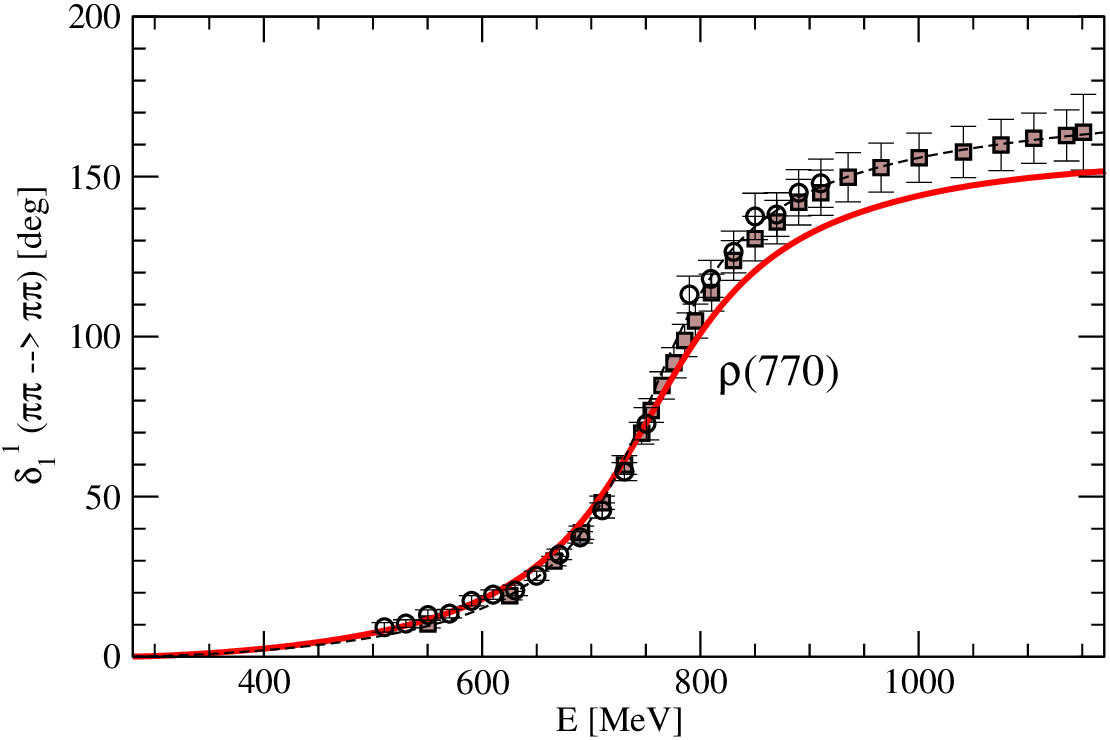} \\
\end{center}
\caption{Solution for the meson-meson interaction [(red) solid lines]. For comparison, the
results of ref.~\cite{Oller:1998hw} are shown with the dashed lines. Data: for $\delta_0^0$:
(pink) diamonds~\cite{Froggatt:1977hu}, (blue) squares~\cite{Batley:2007zz}, (black)
circles: average as defined in ref.~\cite{Oller:1998zr}; for $\delta_{1/2}^0$:
circles~\cite{Aston:1987ir}, triangles up: average as defined in ref.~\cite{Oller:1998zr};
for $\delta_{3/2}^0$: circles~\cite{Linglin:1973ci}, triangles up~\cite{Estabrooks:1977xe};
for $\delta_2^0$: crosses~\cite{Rosselet:1976pu}, squares~\cite{Schenk:1991xe}, triangles
up~\cite{Janssen:1994wn}; for $\delta_{1/2}^1$: triangles up~\cite{Mercer:1971kn},
circles~\cite{Estabrooks:1977xe}; for $\delta_1^1$: squares~\cite{Estabrooks:1974vu},
circles~\cite{Lindenbaum:1991tq}.
}
\label{fig:phase_shifts}
\end{figure}

The meson-meson amplitude in the formulation of ref.~\cite{Oller:1998hw} is
\be
T=V_2\left(V_2-V_4-V_2\,G\,V_2\right)^{-1}V_2
\label{ia}
\ee
with partial-wave projected transition potentials $V_2$ and $V_4$ from the LO and NLO chiral
Lagrangian.  The meson-meson loop function $G$ is factorized from the  potentials $V$
(on-shell reduction), so that $V$ and $G$ in eq.~(\ref{ia}) are simply matrices in channel
space. The expressions for the $V$ can be found in ref.~\cite{Oller:1998hw}~\footnote{
There is a typing error in the denominator of eq.~(B16) of ref.~\cite{Oller:1998hw} that should be $3F_\pi^2F_\eta^2$.
}
 or easily obtained
from there using crossing symmetry, see, e.g., ref.~\cite{Guo:2011pa}. After partial wave
projection, the various particle channels are combined to the possible quantum numbers in
meson-meson scattering.  The normalization of $T$ in eq.~(\ref{ia}) is such that  $\tan
2\delta=-{\rm Re}\,T/({\rm Im}\,T+4\pi E/q_{\rm c.m.})$ with $\delta$ the phase shift and $q_{\rm c.m.}$ the momentum of
the particles in the center-of-mass frame. See refs.~\cite{Oller:1998hw,Oller:1997ti} for the
two-channel case. Fitting the low-energy constants appearing in $V_4$, a good description of
the data can be obtained. We have reproduced the result of ref.~\cite{Oller:1998hw} as shown
in figure~\ref{fig:phase_shifts} by the dashed lines. For the calculation, we use the meson
decay constants $F_\pi=~92.4$~MeV, and, from ref.~\cite{Gasser:1984gg}, $F_K=1.22\,F_\pi$ and
$F_\eta=1.3\,F_\pi$ to be able to compare with the results of ref.~\cite{Oller:1998hw}. As we
have tested, changes from using more modern values of $F_K=(1.192\pm
0.007)\,F_\pi$~\cite{Bernard:2007tk} and $F_\eta=118.4\pm 8.0$~MeV~\cite{Kampf:2009tk} 
can be absorbed in the low-energy constants without
modifying the result significantly.

The agreement of the solution of ref.~\cite{Oller:1998hw} (dashed lines in
fig.~\ref{fig:phase_shifts}) with the phase shift data is remarkable, except for the
combination of isospin, angular momentum, strangeness $(I,L,S)=(1/2,0,-1)$, i.e. precisely
the quantum numbers of the $\kappa(800)$ we are interested in. Thus, we have performed a
refit to the data shown in figure~\ref{fig:phase_shifts} (i.e. from the respective
thresholds up to $E=1.2$~GeV), with details given in the following: the $\kappa(800)$
channel is included in the $\chi^2$ with a special weight of 7. The pole position $z_0$ of
the $\sigma(600)$ in the $(I,L,S)=(0,0,0)$ channel has been determined very precisely and
model-independently in ref.~\cite{Caprini:2003ta}, $z_0=(441\pm i\,272)$~MeV. Thus, we
simply consider this value as an additional data point that is included in the $\chi^2$,
with an artificial error of $|\Delta z_0|=30$~MeV. While the data in the $\kappa$ channel
are included with an increased weight, the $\chi^2$'s from the repulsive $I=2$ $\pi\pi$ and
the $I=3/2$ $\pi K$ channels are included with a weight of $0.2$, while the data from the
$\rho$ channel enter only with a weight of $0.1$ (note the data in that channel are quite
precise). The value of the low-energy constant $L_3$ is particularly sensitive to the
$\rho(770)$-channel and has been fixed in the fit to reproduce the bulk features of
this resonance. With these trade-offs to achieve a better data description in the
$\kappa(800)$ channel, the fit was performed, paying special attention to avoid spurious
poles in the solution --- the inverse amplitude method in the form used here is, to some
extent, prone to generating narrow poles that are not physical. An example, from the
solution of ref.~\cite{Oller:1998hw}, is visible in the dashed line in
figure~\ref{fig:phase_shifts}, for the $\delta_{1/2}^1$ phase at higher energies. In the
present fit, such solutions were discarded. 

\begin{figure}
\begin{center}
\includegraphics[width=0.32\textwidth]{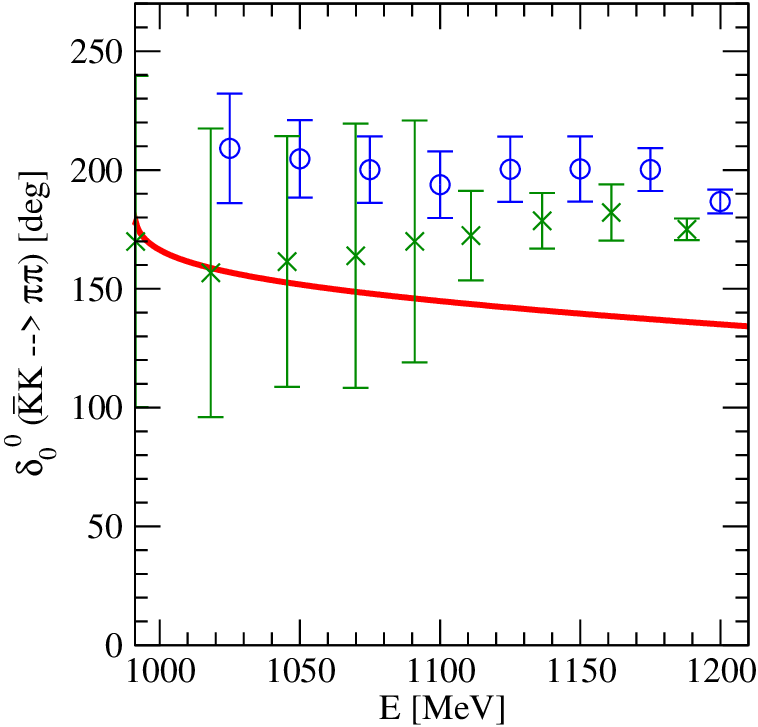}
\includegraphics[width=0.32\textwidth]{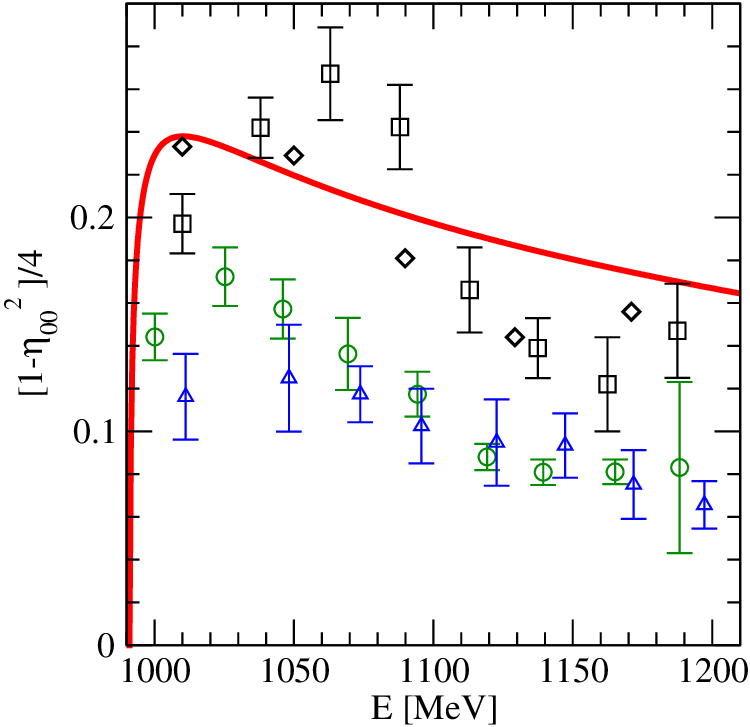}
\includegraphics[width=0.32\textwidth]{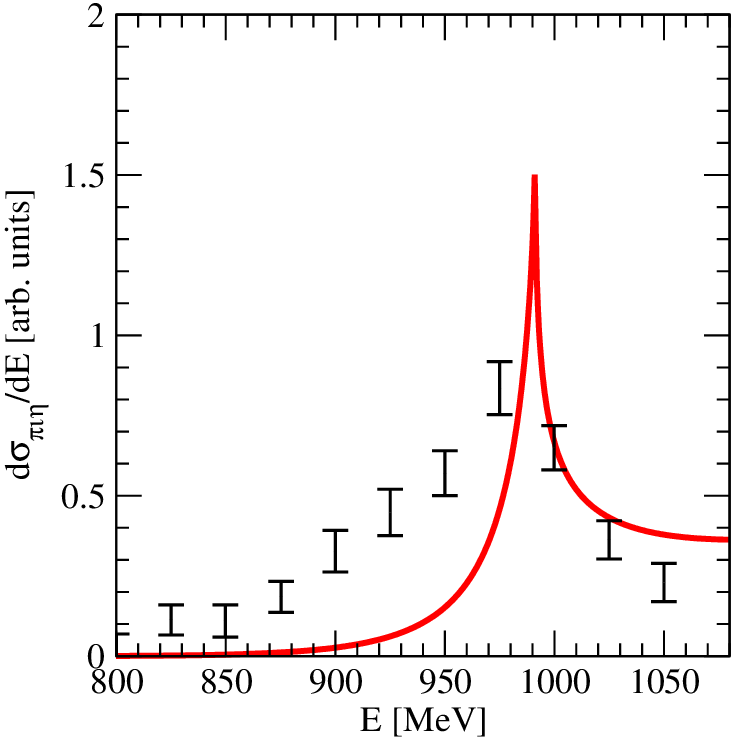}
\end{center}
\caption{Further results (red (solid) lines; data not included in $\chi^2$) for the
$(ILS)=(0,0,0)$ sector [$f_0(980)$ meson] (left and center panel) and for the $\pi\eta$
distribution [$a_0(980)$ meson] (right panel). Data: left and center:
diamonds~\cite{Froggatt:1977hu}, crosses (X)~\cite{Martin:1979gm},
circles~\cite{Cohen:1980cq}, triangles~\cite{Etkin:1981sg},
squares~\cite{Lindenbaum:1991tq}; right: ref.~\cite{Armstrong:1991rg}, background from there
incoherently subtracted, see ref.~\cite{Oller:1998zr}.  }
\label{fig:more_phase}
\end{figure}

The result of the fit is shown by the (red) thick solid lines in
figure~\ref{fig:phase_shifts}. In figure~\ref{fig:more_phase} the performance of the present
fit is compared to additionally available data that were, however, not included in the
$\chi^2$.
Part of the data enter the $\chi^2$ with a weight different from one, and there is not much
sense in calculating the reduced $\chi^2$, but the quality of the fit is obviously good, in
particular in the $\kappa(800)$ channel. The trade-off paid for this agreement, visible in
the $\rho (770)$ and the repulsive $\pi\pi$, $\pi K$ channels, is relatively small. 

The values for the low-energy constants and $q_{\rm max}$ (the cut-off in the loop function
$G$~\cite{Oller:1998hw}) are shown in table~\ref{tab:fit_parameters}. These numbers are
similar to those of ref.~\cite{Oller:1998hw} and also of the same order as they arise in
CHPT, although there is no need for them to be precisely equal to a standard CHPT
calculation due to a different fitted data range and conceptual differences to a full ${\cal
O}(p^4)$ calculation.
\begin{table}
\begin{center}
\begin{tabular}{llll}
\hline \hline
$L_1$				& $L_2$				& $L_3$				& $L_4$ 			\bigstrut[t]\\ 
$0.873^{+0.017}_{-0.028}$	& $0.627^{+0.028}_{-0.014}$	& $-3.5$ [fixed]		& $-0.710^{+0.022}_{-0.026}$	\bigstrut[b]\\
\hline
$L_5$				& $L_6+L_8$			& $L_7$				& $q_{\rm max}$~[MeV] 		\bigstrut[t]\\	
 $2.937^{+0.048}_{-0.094}$	& $1.386^{+0.026}_{-0.050}$	& $0.749^{+0.106}_{-0.074}$	& $981$ [fixed] 		\bigstrut[b]\\
\hline \hline
\end{tabular}
\end{center}
\caption{Fitted values for the $L_i$~[$\times 10^{-3}$] and $q_{\rm max}$~[MeV] with their
uncertainties. As mentioned in the text, $L_3$ and $q_{\rm max}$ have been fixed in the
determination of the error.} 
\label{tab:fit_parameters} 
\end{table}
In addition, the non-linear parameter errors of the fit were determined and are also quoted
in table~\ref{tab:fit_parameters}. These errors are only indicative, because there are data
in the fit that were included in the $\chi^2$ with weights different from one; the
parameter errors are probably considerably larger than those quoted in
table~\ref{tab:fit_parameters}. Also, in the determination of the error, $L_3$ and $q_{\rm
max}$ needed to be fixed. Otherwise, too large parameter correlation would have inhibited
the determination of the error using the usual $\chi^2_{\rm best}+1$-criterion, where
$\chi^2_{\rm best}$ is the $\chi^2$ at its minimum. Apart from this, the errors indicate
that the low-energy constants can be well constrained from the data.

The resulting pole positions and residues are shown in table~\ref{tab:popores}.
\begin{table}
\begin{center}
\noindent\begin{tabular}{rrrllr@{}lr@{}lr@{}l}
\hline \hline
$I$	& $L$	& $S$	& Resonance	& sheet	& \multicolumn{2}{c}{$z_0$~[MeV]}& \multicolumn{2}{c}{$a_{-1}$~[$M_\pi$]}& \multicolumn{2}{c}{$a_{-1}$~[$M_\pi$]}\bigstrut\\
\hline
$0$	& $0$	& $0$	& $\sigma(600)$	& $pu$	& $434$	& $+i\,261$		& $-31$	& $-i\,19\,(\bar KK)$		& $-30$	& $+i\,86\,(\pi\pi)$		\bigstrut[t]\\
	&	&	& 		& $uu$	& $[411$& $+i\,233]$		&$[0.2$	&$-i\,0.2\,(\bar KK)]$		& $[-29$& $+i\,42\,(\pi\pi)]$		\\
$0$	& $0$	& $0$	& $f_0(980)$	& $pu$	& $1003$& $+i\,15$		& $16$	& $-i\,79\,(\bar KK)$		& $-12$	& $+i\,4\,(\pi\pi)$		\\
$1/2$	& $0$	& $-1$	& $\kappa(800)$	& $pu$	& $815$	& $+i\,226$		& $-36$	& $+i\,39\,(\eta K)$		& $-30$	& $+i\,57\,(\pi K)$		\\
	&	&	& 		& $uu$	& $[804$& $+i\,103]$		&$[-0.4$& $+i\,15\,(\eta K)]$		& $[-0.1$&$+i\,34\,(\pi K)]$		\\
$1$	& $0$	& $0$	& $a_0(980)$	& $pu$	& $1019$& $-i\,4$		& $-10$	& $-i\,107\,(\bar KK)$		& $21$	& $-i\,31\,(\pi\eta)$		\\
$0$	& $1$	& $0$	& $\phi(1020)$	& $p$	& $976$	& $+i\,0$		& $-2$	& $+i\,0\,(\bar KK)$		& ---	& 				\\
$1/2$	& $1$	& $-1$	& $K^*(892)$	& $pu$	& $889$	& $+i\,25$		& $-10$	& $+i\,0.1\,(\eta K)$		& $14$	& $+i\,4\,(\pi K)$		\\
	&	&	& 		& $uu$	& $[915$ & $+i\,22]$		& $[-6.4$&$+i\,0.1\,(\eta K)]$		&$[12$& $+i\,3\,(\pi K)]$		\\ 	
$1$	& $1$	& $0$	& $\rho(770)$	& $pu$	& $755$	& $+i\,95$		& $-11$	& $+i\,2\,(\bar KK)$		& $33$	& $+i\,17\,(\pi\pi)$		\\
	&	&	& 		& $uu$	& $[791$& $+i\,80]$		& $[-7$	& $+i\,1\,(\bar KK)]$		& $[27$	& $+i\,13\,(\pi\pi)]$		\bigstrut[b]\\
\hline \hline
\end{tabular}
\end{center}
\caption{Pole positions $z_0$~[MeV] and residues $a_{-1}[M_\pi]$ in the global fit to
meson-meson observables. $I,\,L,\,S$ denote isospin, angular momentum, and strangeness,
respectively. For the sheets, e.g. {\it pu} means physical (unphysical) sheet for the
channel with higher (lower) threshold, in order. Values in brackets refer to poles that
should not be compared to PDG values, because they are on hidden sheets. For the residues,
e.g. $(\pi K)$ means the residue of $T(E)$ of eq.~(\ref{ia}), of the $\pi K\to \pi K$
transition. Note that for the poles in the lower half plane, $a_{-1}(z_0^*)=a_{-1}^*(z_0)$.}
\label{tab:popores}
\end{table}
If there is more than one entry for a resonance, the one without brackets corresponds to the
standard sheet that is considered for the resonance. For example, the $\sigma(600)$
resonance is on the physical sheet with respect to the heavier $\bar KK$
channel, but on the unphysical one with respect to the lighter $\pi\pi$ channel,
i.e. on the $pu$ sheet. The $uu$ sheet with the other pole of the $\sigma$ is
far away from the physical axis at the $\sigma$ energies. For a general discussion of the
analytic structure in hadron-hadron scattering, see, e.g., ref.~\cite{Doring:2009yv}. The
$a_0(980)$ is special because its pole is on the $(pu)$ sheet, but at an energy above the
$\bar KK$ threshold. This means the closest point of the physical axis to this pole is the
$\bar KK$ threshold itself, which indeed shows a pronounced cusp in the $\pi\eta$
invariant mass spectrum as shown in figure~\ref{fig:more_phase}. In the solutions of
refs.~\cite{Oller:1997ti,Oller:1998hw} the situation of the $a_0(980)$ pole is qualitatively
the same. The role of threshold cusps and the possibility to disentangle their structure
from lattice levels has been discussed in ref.~\cite{Doring:2011vk}.  

The pole position of the $\sigma(600)$ shown in table~\ref{tab:popores} is close to the
value extracted in the dispersive analysis of ref.~\cite{Caprini:2003ta} which is not a
surprise as that pole position is included in the $\chi^2$. The pole position of the
$K^*(892)$ is very close to the value quoted in PDG~\cite{Nakamura:2010zzi}. This is
important as the main objective of this study is the $(\eta K,\,\pi K)$ system.

We would finally like to stress that in the presently considered UCHPT approach the
considered channel space is limited to the two-body channels of scalar mesons. Multi-meson
states or channels with (axial) vector particles are omitted, mostly because there is no
phenomenological requirement for their inclusion at the energies of interest. The finite
volume structure of multi-meson states is considerably more complicated than the two-body
case considered here, and beyond the scope of this study.


\subsection{Predicted lattice levels in the meson-meson sector}
\label{sec:predictions}
The discretization of the loop integral $G$ in eq.~(\ref{ia}) generates the finite volume
effects. In ref.~\cite{Doring:2011vk}, the discretized meson-meson loop function for
$S$-waves has been derived with the result
\ba
\tilde G_{j}&=&\tilde G_{S,j}+\left(G_{j}-G_{S,j}\right) \ , \non
\tilde G_{S,j}&=&\frac{1}{L^3}\sum_{\vec n}\,f_j(|\vec q|)\,F(q)\ , \quad
G_{j}= \int\limits^{|\vec q_{\rm max}|}\frac{d^3\vec q}{(2\pi)^3}\,f_j(|\vec q|) \ , \quad
G_{S,j}=\int\frac{d^3\vec q}{(2\pi)^3}\,f_j(|\vec q|)\,F(q) \ ,
\non 
\vec q&=&\frac{2\pi}{L}\,\vec n, \,\,\, \vec n\in \mathds{Z}^3 \ ,
\quad
f_j(q)=\frac{1}{2\omega_{j,1}(q)\,\omega_{j,2}(q)}\,\,
\frac{\omega_{j,1}(q)+\omega_{j,2}(q)}{E^2-(\omega_{j,1}(q)+\omega_{j,2}(q))^2 [+i\epsilon]}
\label{tildegs}
\ea
with $\omega_{j,i}=\sqrt{M_i^2+|\vec q|^2}$, $M_i$ the masses of the two mesons in channel
$j$, and $F$ is a form factor introduced in ref.~\cite{Doring:2011vk} to avoid artefacts
($e^{-M_\pi\,L}$ effects) from the sharp cut-off at $\vec q_{\rm max}$. The function $\tilde
G_j$ is independent of this form factor, and $\tilde G_{S,j}-G_{S,j}$ can be rewritten in
terms of the ${\cal Z}_{00}$ function~\cite{Doring:2011vk} up to relativistic effects that are
of order $e^{-M_\pi\,L}$. As has been studied in ref.~\cite{Doring:2011vk}, the relativistic
effects provided by eq.~(\ref{tildegs}) can become significant. Up to the same $e^{-M_\pi\,L}$
effects the present discretization procedure is equivalent to the $K$-matrix formulation
derived in ref.~\cite{Bernard:2010fp}.  Note also that in the extraction of the scattering
phase, in the one-channel case only the combination $\tilde G_{j}-G_{j}$ enters that is
independent of the used form factor and cut-off at $\vec q_{\rm max}$~\cite{Doring:2011vk}.

The lattice levels are obtained from the singularities of eq.~(\ref{ia}) with the
discretized $\tilde G_j$,
\be
\det|V_2-V_4-V_2\,\tilde G\,V_2|=0,\quad \tilde G={\rm diag}\,(\tilde G_{1},\tilde G_{2})
\label{det}
\ee
where $\tilde G$ is a diagonal matrix in channel space that has two dimensions for the
cases $(I,L,S)=(0,0,0),\,(1,1,0)$ [$\bar KK$, $\pi\pi$ channels],
$(I,L,S)=(1/2,0,-1),\,(1/2,1,-1)$ [$\eta K$, $\pi K$ channels], and for $(I,L,S)=(1,0,0)$
[$\bar KK$, $\pi\eta$ channels]. Only one channel is present for the cases
$(I,L,S)=(3/2,0,-1)$ [$\pi K$ channel], $(I,L,S)=(2,0,0)$ [$\pi\pi$ channel], and
$(I,L,S)=(0,1,0)$ [$\bar KK$ channel], in which case eq.~(\ref{det}) is one-dimensional with
$\tilde G=\tilde G_j$.

For the $P$-wave states, one has to complement eq.~(\ref{tildegs}) with the corresponding
angular momentum structure for the vertices, as spherical symmetry is broken on the lattice.
In the general case of particles with spin $s_1,\,s_2$ coupling to a total spin $S$, the
angular structure $\gamma$ of a vertex that couples spin $(S,m)$ and orbital angular
momentum $(\ell,M-m)$ to a total $(J,M)$ is
\be
\gamma_\alsub={\cal C}(S,\ell,J;m,M-m,M)\,Y_{\ell,m-M}(\theta,\phi)\,(-1)^{M-m}
\non
\label{gamma}
\ee
with the Clebsch-Gordan coefficients ${\cal C}$ and
$\boldsymbol{\alpha}=(J,M,\ell,S,m,\theta,\phi)$ abbreviates the quantum numbers and polar
and azimuthal angles $\theta\in [0,\pi]$ and $\phi\in [0,2\pi]$. Note that the angles depend
on the discretized momenta,
\be
\theta=\arccos\frac{n_z}{\sqrt{n_x^2+n_y^2+n_z^2}},\quad \phi=\arctan (n_y/n_x) 
\label{angles}
\ee 
with $\vec n=(n_x,n_y,n_z)\in \mathds{Z}^3$.
For the present case of spinless meson-meson scattering in $P$-wave, 
\be
\gamma^*_\alsub\gamma_\alpsub=3\,\cos^2\theta
\ee
and, thus, eq.~(\ref{tildegs}) becomes
\be
\tilde G_{S,j}^{\rm P-wave}=\sum_{\vec n}\,\gamma^*_\alsub\gamma_\alpsub 
\,f_j(|\vec q|)\,F(q)=\tilde G_{S,j} \ , 
{\rm i.e.}, \ \tilde G_{j}^{\rm P-wave}=\tilde G_{j} \ .
\ee
The equation also indicates that for the special case of $P$-wave, the $\tilde G_{j}^{\rm
P-wave}$ function is the same as for $S$-wave, $\tilde G_{j}$, given in 
eq.~(\ref{tildegs})~\footnote{
This may be seen as follows: in $S$-wave, a contribution to the multiplicity at fixed $m$ can
be written as $m=n_1^2+n_2^2+n_3^2$. The multiplicity for a set of such $n_i$ is
$2^{(3-N_z)}\,3!/N_g!$ where $N_z$ is the number of zeros in $(n_1,n_2,n_3)$ and $N_g$ the
number of equal $n_i$. On the other hand, for $P$-wave,
$3\cos^2\theta=3n_3^2/(n_1^2+n_2^2+n_3^2)$. Weighting each distinguishable permutation of
$(n_1,n_2,n_3)$ with this factor, it is easy to see that the multiplicity at fixed $m$ is the
same as for the $S$-wave case.
}.
An analog situation has been found for the mass shift of bound states in a box in
refs.~\cite{Konig:2011nz,Koenig:2011ti}. This is no longer the case for $D$- and higher
waves. Note also that sums $\sum_{\vec n}$ with an $S$-wave and a $P$-wave vertex disappear,
i.e. there is no mixing of $S$- and $P$-waves. As is well known, there is mixing of $S$- and
$G$-waves, and of $P$- and $F$-waves, but this is not an issue in the present study because
the respective higher partial waves are small and, anyway, not generated from the NLO-terms
used in this study.

\begin{figure}
\begin{center}
\includegraphics[width=0.95\textwidth]{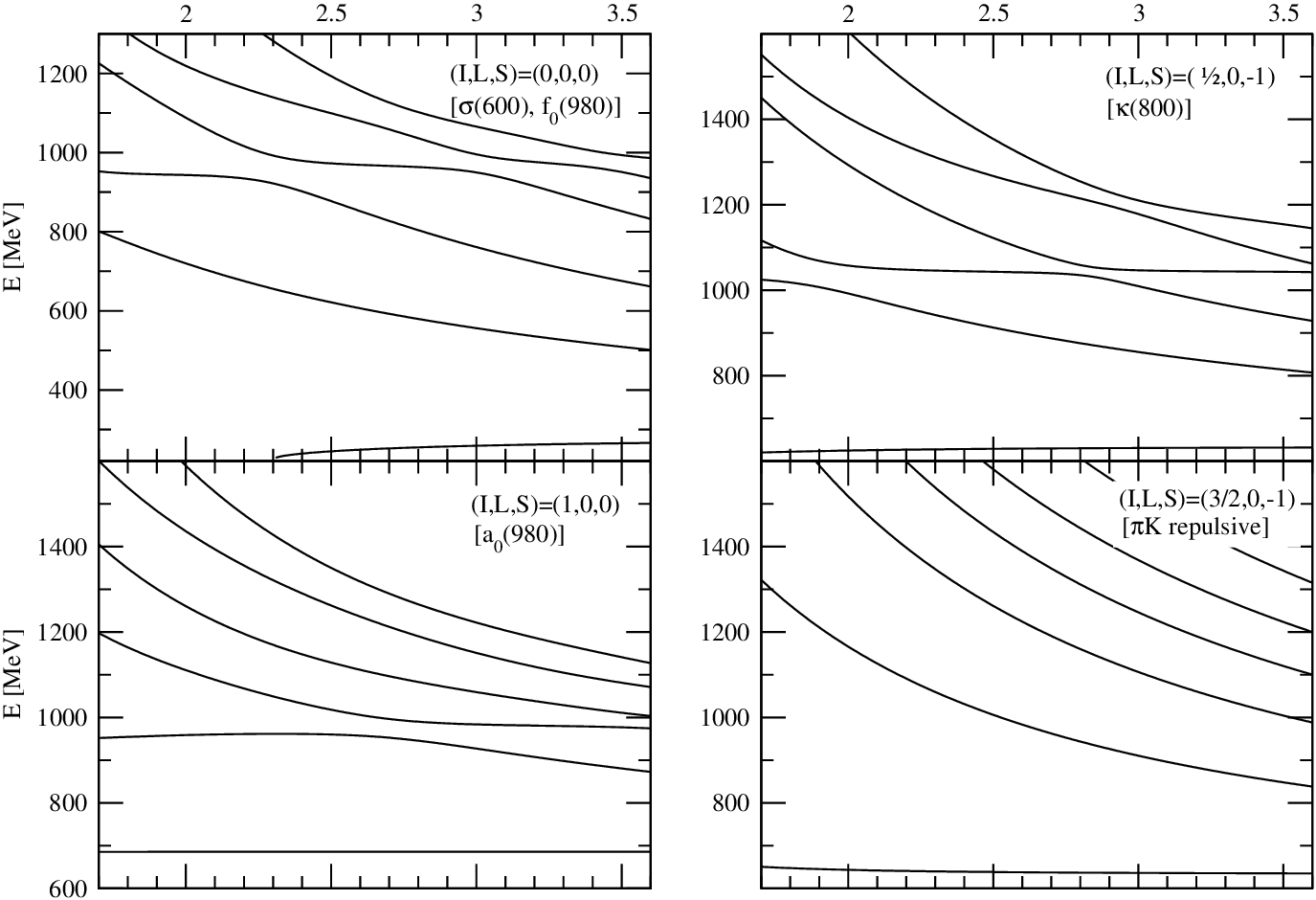}\\
\vspace*{-0.2cm}
\includegraphics[width=0.95\textwidth]{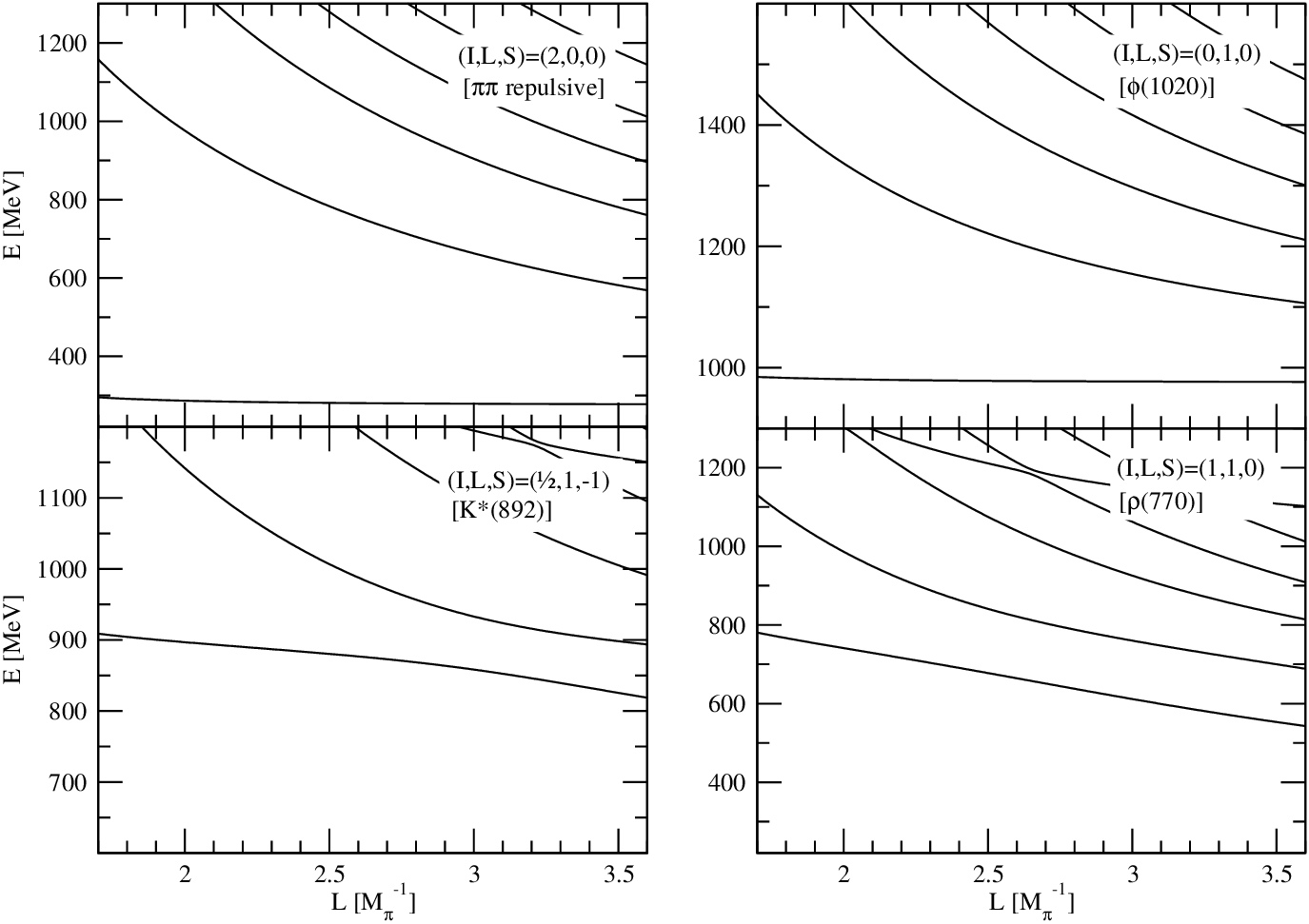}
\end{center}
\caption{The first few levels for the considered reactions as calculated from the present
solution for meson-meson scattering. Also, the corresponding resonances are indicated.
}
\label{fig:levels}
\end{figure}

With the $\tilde G$ function derived for $S$- and $P$-waves, and using the LO potentials
$V_2$ and NLO potentials $V_4$, with the low-energy constants of the fit of
table~\ref{tab:fit_parameters}, eq.~(\ref{det}) can now be solved for its tower of zeros and
at different box sizes $L$. The results are shown in figure~\ref{fig:levels} for the eight
considered reactions.
For the $(I,L,S)=(0,0,0)$ case, one can perceive the avoided level crossing around
$E=990$~MeV that comes from a combined effect of the $\bar KK$ threshold and the $f_0(980)$
resonance located there. The situation of resonances close to thresholds and methods to
extract those resonances from lattice data, e.g. by using twisted boundary conditions, have
been discussed in refs.~\cite{Bernard:2010fp,Doring:2011vk}. A similar situation is also
given for the $a_0(980)$ [cf. figure~\ref{fig:levels}] although in this case the resonance
pole is considerably hidden on the $pu$ sheet in the present solution, see
table~\ref{tab:popores}. Also in case of the $\kappa(800)$ quantum numbers, shown to the
upper right in figure~\ref{fig:levels}, avoided level crossing is visible, coming in this
case solely from the $\eta K$ threshold, as the $\kappa$ resonance pole is at lower
energies. For the $P$-wave cases of the $K^*(892)$ and $\rho(770)$ quantum numbers, avoided
level crossing is visible at higher energies $E>1.1$~MeV. This is due to the first free
level of the higher channel, i.e. $\eta K$ for the  $K^*(892)$ and $\bar KK$ for the
$\rho(770)$ quantum numbers. As these channels couple very weakly to the dominant $\pi K$ and
$\pi\pi$ channels, respectively, the crossing is almost ideal, i.e. the different levels
come very close at the crossing energy. Note that, in contrast to $S$-wave channels, there
is no avoided level crossing at the respective higher thresholds for the $P$-wave reactions,
nor is there a lowest level close to the lower thresholds. An exception is given for the
$\phi$-channel, but there the level at $E\sim 980$~MeV arises from the $\phi(1020)$
resonance that is just below the $\bar KK$ threshold, cf. table~\ref{tab:popores}. The
explanation for these findings is postponed until section~\ref{sec:lowest}.

For the $\kappa(800)$ channel, around the energy of the resonance, there is no structure
visible in the levels shown in figure~\ref{fig:levels}, and the same applies to the
$\sigma(600)$. For the latter case, already studied in ref.~\cite{Doring:2011vk}, it is
known that the $\sigma(600)$ pole can still be reconstructed, even if its position far in
the complex plane results in a totally washed-out avoided level crossing. The situation
should be similar for the $\kappa(800)$, and this question is addressed in
section~\ref{sec:results}.

It should be noted that in the present approach we consider the finite volume effects on the
meson-meson interaction, but do not attempt to determine finite size effects on the lattice
levels. A comprehensive treatment of such systematic uncertainties would require to address
the actual lattice action that generates the levels, but that goes far beyond
the scope of the present study.


\section{Reconstructing the \texorpdfstring{$\boldsymbol{\kappa(800)}$}{kappa(800)}}
\label{sec:results}


\subsection{A suitable potential for resonance extraction}
\label{sec:whichpotential}
In the one-channel case, the L\"uscher formalism provides a model-independent connection
between phase shift (in the infinite volume limit) and lattice
levels~\cite{Luscher:1986pf,Luscher:1990ux,Luscher:1991cf}. For the extraction of scattering
lengths, see, e.g., also refs.~\cite{Lage:2009zv,Beane:2003da}. In the two-channel case
(see, e.g., ref.~\cite{Lage:2009zv}), phase shifts, poles and residues can still be
reconstructed model-independently provided three independent measurements of lattice levels
at the same energy, to determine the channel transitions $V_{11}$, $V_{12}$, and $V_{22}$.
This information can be obtained, e.g., from twisted boundary conditions or using spatially
asymmetric boxes, as pointed out in refs.~\cite{Bernard:2010fp,Doring:2011vk}. 

However, if lattice data are provided with finite accuracy  ---present-day lattice simulations
do not provide data for more than a few volumes, and with rather large error bars--- the
reconstruction of the infinite volume limit becomes a challenging task. A possible way of
proceeding is to expand a general scattering potential $V$ in powers of the scattering energy,
in one or more channels. The expansion parameters are then fitted to the lattice levels, and
with the same potential $V$ the infinite volume limit can be recovered. Even broad resonances
like the $\sigma(600)$ or resonances close to a threshold like the $f_0(980)$ can be
reconstructed as shown in ref.~\cite{Doring:2011vk}.

The two-channel fit potential used in ref.~\cite{Doring:2011vk} is given by
\be
V_{ij}=a_{ij}+b_{ij}(s-4M_K^2)\, 
\label{fitv}
\ee
where $i,j=1,2$ indicate the channel. The six parameters $a_{ij}=a_{ji},\,b_{ij}=b_{ji}$
provide an expansion up to terms linear in $s=E^2$, with the expansion point $s_0=4M_K^2$ at
the $\bar KK$ threshold that is suitable for the study of the $f_0(980)$ resonance since the
parameter correlations are minimized. The problem at hand in ref.~\cite{Doring:2011vk} is
specially suited for the choice of the fit potential shown in eq.~(\ref{fitv}), because the
LO chiral interaction in the $(\bar KK,\,\pi\pi)$ system depends only linearly on $s$. If
then pseudo-data, generated from the very interaction, are fitted with the ansatz of
eq.~(\ref{fitv}), ideally a $\chi^2$ of zero can be obtained. In the present case of the
$(\eta K,\,\pi K)$ system, the LO interaction depends already on $s$ and $t$, and the NLO
terms have $s^2$-, $t^2\,\cdots$-dependence.  A possible refinement of the potential $V$
could, thus, be an expansion not only in $s$ but also in $t$,
\be
V_{ij}=a_{ij}+b_{ij}(s-s_0)+b'_{ij}(t-t_0)+c_{ij}(s-s_0)^2+\cdots\, \ .
\label{fitvtry}
\ee
The dependence on the Mandelstam variable $u$ can be expressed in this ansatz by $s$, $t$
and a constant, and terms $\sim (u-u_0)$ should not be included explicitly. One should note
that in $S$-wave, terms proportional to $t$ are to a good approximation linear functions of
$s$, and large correlations between $a,b,b'$ appear. It is then more efficient to absorb the
$t$-dependence in the parameter $b$, and also in $c$ for the small order-$s^2$ differences
between $s$ and $t$. 

As we have just seen, in the presence of terms $\sim t$ and higher order terms from NLO, a
quadratic term of the form $(s-s_0)^2$ is indispensable in the expansion of the fit
potential. This will, of course, introduce additional free parameters in the fit potential.
Together with the fact that only very few data points are available in actual lattice
simulations, there will still be large parameter correlations and, worse, the fit will have
too much freedom in between those data points and generate solutions that have a good
$\chi^2$ but should not be accepted because the natural hierarchy of the expansion is no
longer given ---e.g., the term $\sim (s-s_0)^2$ may be unnaturally large compared to the
terms $(s-s_0)^1$ and $(s-s_0)^0$. The situation can be improved by providing the expansion
the well-defined, model-independent LO term of the chiral expansion. This term almost
saturates the constant and the linear parameters $a$ and $b$, and thus considerably
stabilizes the fit. Remaining contributions to $a$ and $b$ will then be much smaller, and
the above-mentioned, unnatural solutions are automatically avoided.

The question is in which form to cast the resulting fit potential. For a resonance analysis,
the inverse amplitude method itself is particularly suited because the organization of the
LO and NLO potentials, $V_2$ and $V_4$, allows for solutions in which resonance terms can be
expressed in terms of the $V_2$, $V_4$~\cite{Oller:1998hw}. Thus, this form of the
parameterization is quite flexible, and we choose it for the final form of the fit potential
used throughout this study. The finally adopted form of the fit potential reads in the
one-channel case
\be
V^{\rm fit}=\left(\frac{V_2-V_4^{\rm fit}}{V_2^2}\right)^{-1},\quad 
V_4^{\rm fit}=a+b(s-s_0)+c(s-s_0)^2+d(s-s_0)^3+\cdots
\label{vfit}
\ee
with $V_2\equiv V_{\rm LO}$ the fixed LO term of the chiral expansion.  In other words, we
take the form of the inverse amplitude, leaving the LO term $V_2$ as it is, and expand $V_4$
in powers of $s$. The $T$ matrix in the infinite volume limit is then obtained as 
\be
T=\frac{1}{(V^{\rm fit})^{-1}-G}, \quad {\rm i.e.,} 
\quad T=V^{\rm fit}+ V^{\rm fit}\, G\, T~,
\label{tv}
\ee
and the lattice levels are given through the condition
\be
(V^{\rm fit})^{-1}-\tilde G=0\quad\leftrightarrow\quad V_2-V_4^{\rm fit}-V_2^2\tilde G=0 
\label{levvfit}
\ee
where $\tilde G=\tilde G_{\pi K},\, G= G_{\pi K}$ from eq.~(\ref{tildegs}) for the
$\kappa(800)$ channel.  In the present study, we do not generalize this fit potential to two
channels. The pseudo-data which will be fitted are, of course, generated from the full
two-channel solution obtained in the previous section. The $\kappa (800)$ is situated well
below the $\eta K$ threshold, and the influence from the $\eta K$ channel should be noticed,
at most, as a small perturbation that can be absorbed in  the fit potential. Note in this
context that the pseudo-phase~\cite{Bernard:2010fp} in the $(\bar KK,\,\pi\pi)$ system is
very close to the true phase up to a few MeV below the $\bar KK$
threshold~\cite{Doring:2011vk}. In any case, these arguments will be tested in the
following: if no good fits could be obtained, the one-channel extraction method would lead
to erroneous phase shifts and pole positions. 

As the LO chiral interaction $V_2$ enters the fit potential explicitly to stabilize the
solution, the question arises to which extent extracted pole positions are biased: for
example, $V_2$ alone already generates poles in $S$-wave that can be identified with the
$\sigma(600)$, $f_0(980)$, and $a_0(980)$ resonances~\cite{Oller:1997ti}. This is also the
case for the $\kappa(800)$ resonance. However, note that $V_4^{\rm fit}$ in eq.~(\ref{vfit})
contains also terms $\sim s^0$ and $\sim s^1$ as they appear in $V_2$. Thus, $V^{\rm fit}$
is sufficiently general to modify the fixed LO contribution, and resonances may or may not
appear in the reconstruction process, as dictated by the lattice data themselves. This
provides a much more model-independent approach to the extraction of the finite volume limit
as it would be the case by making assumptions on the resonance propagator as, e.g., in
refs.~\cite{Gockeler:2008kc,Feng:2010es,Lang:2011mn,Fu:2011xw}.

Finally, it is worthwhile noting that the proposed one-channel extraction method is entirely
independent of the used regularization scheme because only the difference $\tilde G_j-G_j$
appears in the connection between the $T$-matrix in the infinite-volume limit and the
lattice levels, eqs.~(\ref{tv}) and (\ref{levvfit}).


\subsection{Extraction of the \texorpdfstring{$\boldsymbol{\kappa(800)}$}{kappa(800)}}
\label{sec:higher_pole}
Fitting pseudo-data with the potential of eq.~(\ref{vfit}) allows to study the propagation
of the uncertainty from the lattice to the quantities of interest, i.e., the phase shift and
the $\kappa(800)$ pole position in the infinite volume limit. To generate the pseudo-data,
we consider the levels as shown in figure~\ref{fig:levels} to the upper right. The possible
pole position is expected around $E\sim 650-850$~MeV, but at the values of $L$ shown in the
figure, which are values that can be covered in actual lattice simulations, there is no
level at these energies. Still, one can choose the nearby level at the $\pi K$ threshold and
the following one, without getting too close to the $\eta K$ threshold. In principle, one
could include also data from the $\eta K$ threshold region, generalizing the fit potential of
eq.~(\ref{vfit}) to two channels. However, in actual lattice simulations only few data will
be available and the number of free parameters in the two-channel potential would be three
times larger, most of them being weakly constrained (the ones from the $V_{\eta K\to \eta K}$
and $V_{\pi K\to \eta K}$ transitions). This would, thus, be a poor strategy to study the
$\kappa(800)$.

\begin{figure}
\begin{center}
\includegraphics[width=0.40\textwidth]{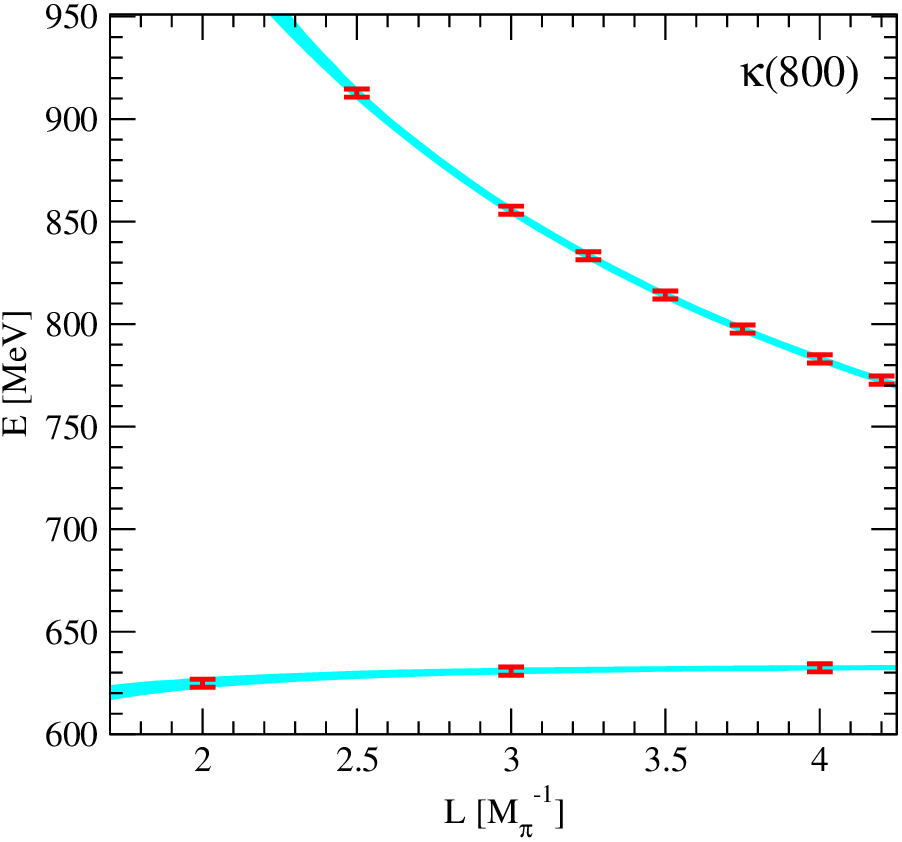} \hspace*{0.3cm}
\includegraphics[width=0.56\textwidth]{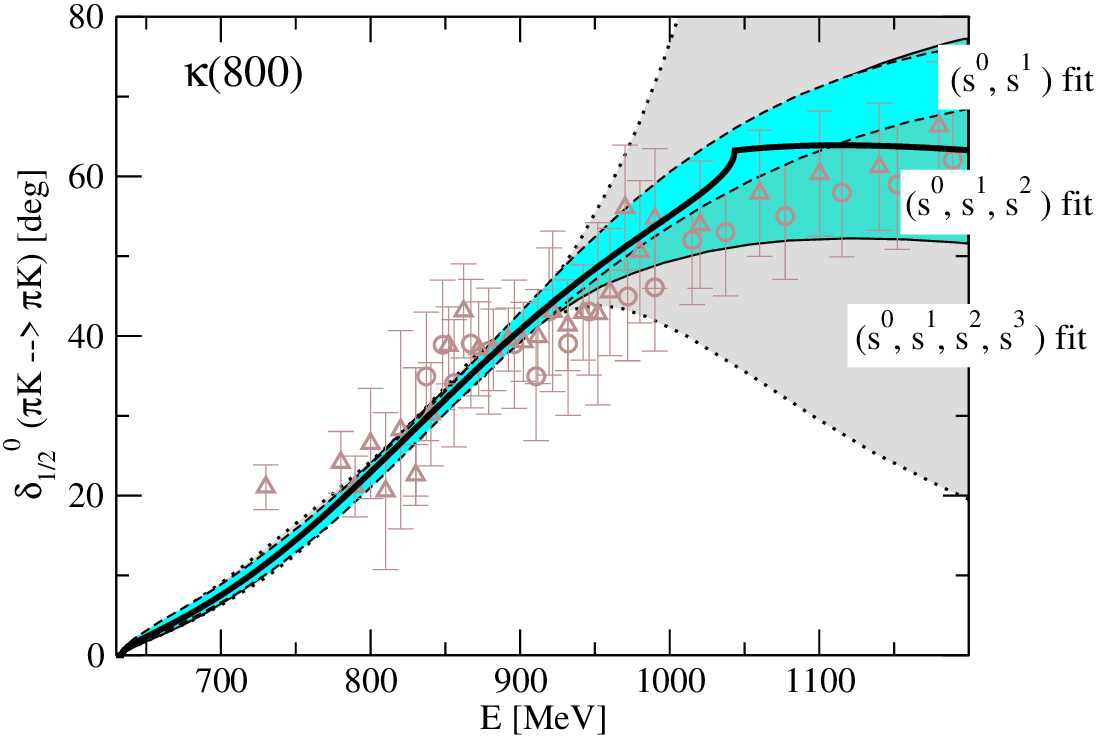}
\end{center}
\caption{Left: Pseudo lattice-data and ($s^0,\,s^1,\,s^2$) fit to those data with
uncertainties (bands). Right: Solid line: Phase shift of the solution that was used to
generate these pseudo lattice-data. Also, the resulting error bands of the $(s^0,\,s^1)$,
$(s^0,\,s^1,\,s^2)$, and $(s^0,\,s^1,\,s^2,\,s^3)$ fits to these data are shown. }
\label{fig:fit_high_pole}
\end{figure}

Given these arguments, the finally chosen pseudo-data from the levels of
figure~\ref{fig:levels} are shown in figure~\ref{fig:fit_high_pole} to the left. 
A constant error of $2$~MeV is assigned to each data point, but we will also study the case
of a 5~MeV error. In ref.~\cite{Doring:2011vk} the additional error, induced by the
statistical displacement of the centroids of the error bars, has been studied. Displacing
the centroid according to a Gauss distribution with half the width of the error bars
themselves, the uncertainties of extracted phase shifts and pole positions were found to be
enlarged by $\sim 40$\%. We will not repeat this exercise in the analyses of this work, but
one should keep in mind additional errors of this size on the extracted quantities.
 
Using eqs.~(\ref{vfit}) in eq.~(\ref{levvfit}), the fit is performed at an expansion point
of $s_0=~(750\,{\rm MeV})^2$ in eq.~(\ref{vfit}). Depending on the power to which the fit
potential is expanded, the fits are labeled $(s^0), \cdots,\, (s^0,\,s^1,\,s^2,\,s^3)$ in
the following. Error bands are generated by accepting all solutions for which $\chi^2\leq
\chi_{\rm best}^2+1$ where $\chi^2_{\rm best}$ is the $\chi^2$ of the best fit. The error
bands for the levels themselves, from the ($s^0,\,s^1,\,s^2$) fit, are shown in
figure~\ref{fig:fit_high_pole} in the left panel, and for the phase shift to the right.
Also, to the right, the solid line shows the phase shift of the solution that was used to
generate the pseudo lattice-data, which lies, as expected, in the center of the error bands
up to $E\sim 900$~MeV while it deviates at higher energies. This is because, first, no
lattice data are fitted beyond $E\sim 920$~MeV, and second, the solution to generate the
data is formulated in the full two-channel approach while the fit potential is in one
channel only. This manifests itself in a cusp of the full solution at the $\eta K$ threshold,
while the extracted phase shifts are smooth there. As the figure shows, fits with higher
power in the fit potential show a larger spread of possible phase shifts at higher energies.
This is expected because more free parameters in the fit potential lead to a better fit of
the level, but also and unavoidably, to a wider error band.

\begin{figure}
\begin{center}
\includegraphics[width=0.8\textwidth]{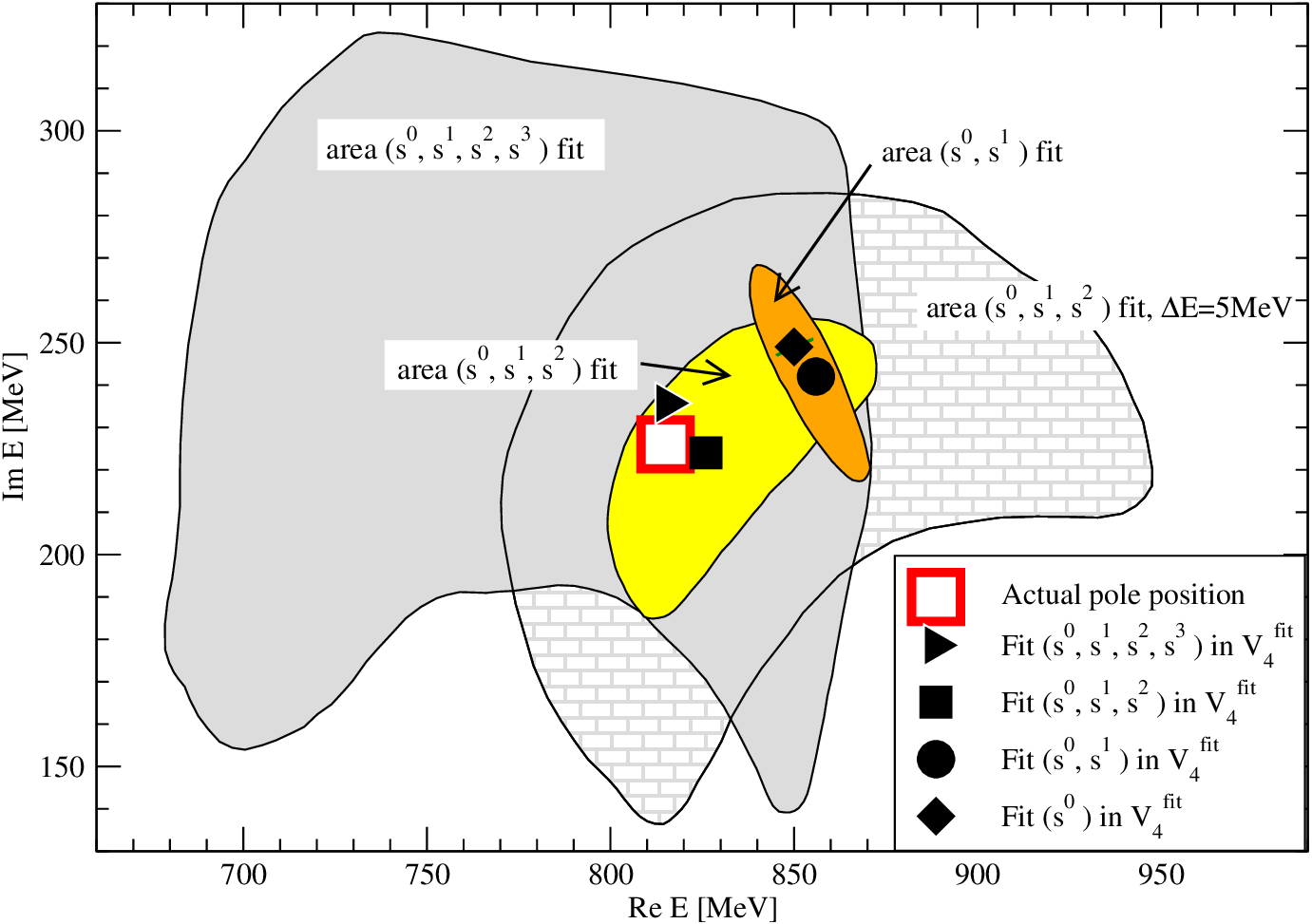}
\end{center}
\caption{Pole reconstruction of the $\kappa(800)$. The actual pole position is obtained from
a global fit to meson-meson scattering data, cf. table~\ref{tab:popores}. Shown are the pole
positions of the $\kappa$ with uncertainty areas as obtained from different fits to pseudo
lattice-data. The lattice data carry an error of 2~MeV, except for the case $\Delta E=5$~MeV
as specified in the figure.
}
\label{fig:pole_spread_high_pole}
\end{figure}

The next question is to which precision the pole position of the $\kappa(800)$ can be
determined. The main problem is the large width of this resonance ---small uncertainties on
the real, physical axis at ${\rm Im}~E=0$~MeV result in large uncertainties far in the
complex plane. In figure~\ref{fig:pole_spread_high_pole} the pole positions as extracted
from the pseudo lattice-data are shown for the different fit functions. The central values
are indicated with the symbols while the areas show the corresponding uncertainties.   
As for the central values, the $(s^0,\,s^1,\,s^2)$ and $(s^0,\,s^1,\,s^2,\,s^3)$ fits are
close to the actual pole position, while the $(s^0)$ and $(s^0,\,s^1)$ fits are not. This
indicates that the $\sim s^2$ term in eq.~(\ref{vfit}) is important. As argued in the
previous section, this term bears part of the $t$- and $u$-dependence of the potential, plus
parts of the NLO interaction. Both turn out to important. Considering the uncertainty areas,
we observe the same behavior as for the phase shifts: the areas grow with increasing number
of fit parameters and higher powers in the potential. In particular, for the $(s^0)$ and
$(s^0,\,s^1)$ fits the areas are very small, giving the wrong impression of high precision
but the areas do not cover the actual pole position. In contrast, the uncertainty area
of the $(s^0,\,s^1,\,s^2)$ and $(s^0,\,s^1,\,s^2,\,s^3)$ fits do cover the actual pole
position. 

While in our analysis of pseudo lattice-data the actual pole position is known, in the
analysis of real lattice data it is not. Still, the previous observations lead to the
formulation of a simple fit strategy: 
\begin{enumerate}
\item
The fit potential should be expanded to higher and higher powers
in $s$. 
\item
Once the central value of the pole position does not change any more ---here, given for the
$(s^0,\,s^1,\,s^2,\,s^3)$ fit--- one can assume convergence. 
\item
One can then use the previous fit to determine the uncertainty area ---here given by the
$(s^0,\,s^1,\,s^2)$ fit and the yellow area in the plot. This can be regarded as the final
result of the analysis.
\end{enumerate}
The chosen data uncertainty of $2$~MeV is very small, and we show in
figure~\ref{fig:pole_spread_high_pole} also the uncertainty area for a $5$~MeV error, for
the $(s^0,\,s^1,\,s^2)$ fit. This results in an error of around $\pm 90$ MeV for the real
part of the pole position and around $\pm 75$~MeV for the imaginary part, i.e. $\pm 150$~MeV
for the width. The residues of the extracted $\kappa(800)$ poles and of all other resonances
analyzed in this study can be found in section~\ref{sec:residues}.


\subsection{Case of a low \texorpdfstring{$\boldsymbol{\kappa(800)}$}{kappa} pole}
\label{sec:lowkappa}
The pseudo-data analyzed in the previous section are based on the global fit to meson-meson
scattering data carried out in section~\ref{sec:inifinitev}. The outcome for the real part of
the $\kappa(800)$ pole position of $z_0=(815\pm i\,226)$~MeV is rather at the upper end of the
pole positions quoted in the PDG~\cite{Nakamura:2010zzi}. To cover also low pole positions of
the $\kappa(800)$, we repeat the analysis using a different solution for which we include, in
the $\chi^2$ of the fit, one of the lowest pole positions quoted in the
PDG~\cite{Nakamura:2010zzi}. The pole position is the result of the Roy-Steiner analysis of
ref.~\cite{DescotesGenon:2006uk} with $z_0=(658\pm i\, 278.5)$~MeV. For simplicity, we use
only the $\pi K$ channel for the chiral unitary model and include, besides the $\kappa$ pole
position, only the partial wave data from the $\pi K$ partial waves
$(I,L,S)=(1/2,0,-1),\,(1/2,1,-1)$ up to $E=1.2$~GeV [see figure~\ref{fig:phase_shifts} for the
data and their references]. The $\kappa$ pole position is included with a chosen error of
$35$~MeV in the $\chi^2$. To constrain the low-energy constants, also $\pi\pi$ scattering data
have been included in the $\chi^2$, but only to ensure the $\sigma(600)$ pole is still
present. Furthermore, we have made sure that the fitted combination of low-energy constants
leads to qualitatively correct solutions in the other meson-meson reactions. 

\begin{figure}
\begin{center}
\includegraphics[width=0.47\textwidth]{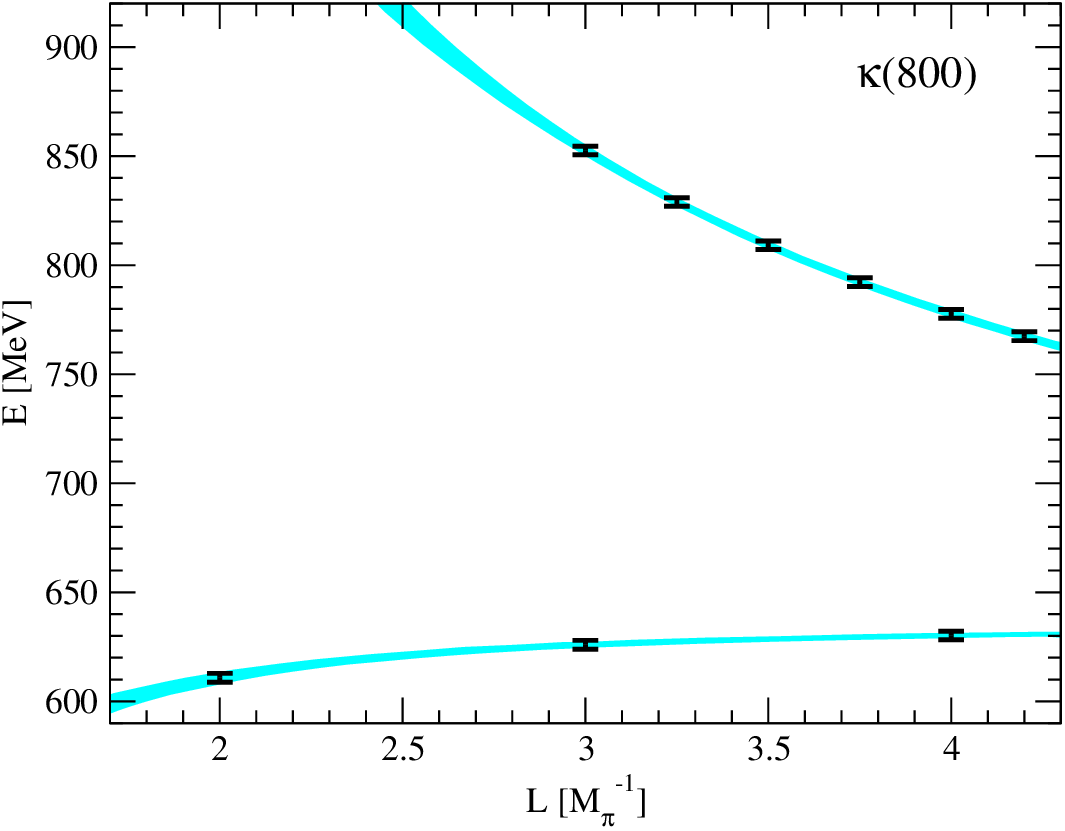} \hspace*{0.3cm}
\includegraphics[width=0.47\textwidth]{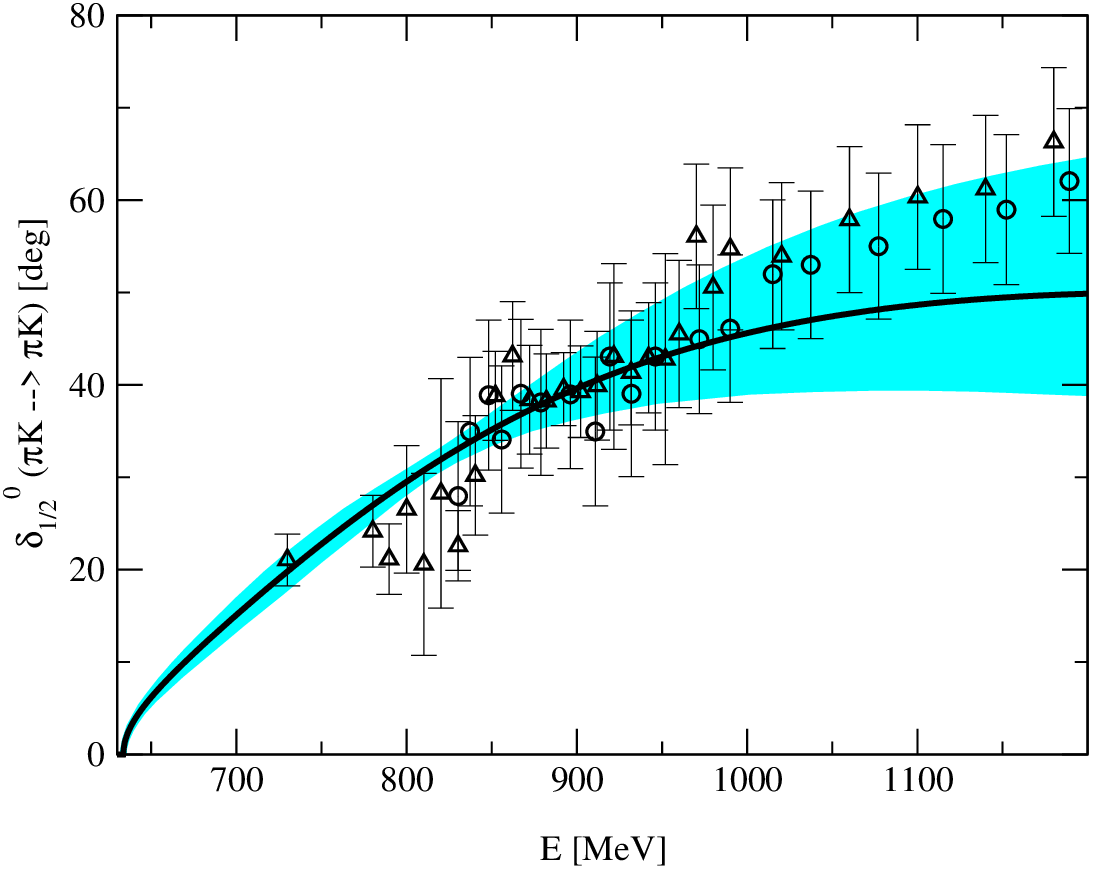}
\end{center}
\caption{Case of a low $\kappa(800)$ pole. Left: Pseudo lattice-data and uncertainty of the
$(s^0,s^1,s^2)$ fit to those data (bands). Right: Solid line: Phase shift for the solution
that was used to generate these pseudo lattice-data. Also, the resulting error band of the
$(s^0,s^1,s^2)$ fit to these data is shown.
}
\label{fig:fit_low}
\end{figure}

The refit of the low-energy constants (except for $L_7$ that does not contribute in 
$\pi K\to~\pi K$) leads to a quantitative description of the data in the $\pi K$ partial waves
$(I,L,S)=(1/2,0,-1),\,(1/2,1,-1)$. This is shown by the solid line in
figure~\ref{fig:fit_low} in the right panel (the $K^*(892)$ channel is not shown again). The
pole position at $z_0=~(679\pm~i\,~271)$~MeV is quite close to the fitted one of
ref.~\cite{DescotesGenon:2006uk} and at the lower end of values quoted in the
PDG~\cite{Nakamura:2010zzi}.

The fit solution provides the pseudo-data that are selected following the same arguments as in
the previous section. The pseudo-data and the error bands of the corresponding
($s^0,\,s^1,\,s^2$) fit are shown in figure~\ref{fig:fit_low} to the left, applying the same
$\chi_{\rm best}^2+1$-criterion as in the previous section. To the right, the error band of
the extracted phase shift from the ($s^0,\,s^1,\,s^2$) fit is indicated with the filled
(cyan) area. 
\begin{figure}
\begin{center}
\includegraphics[width=0.8\textwidth]{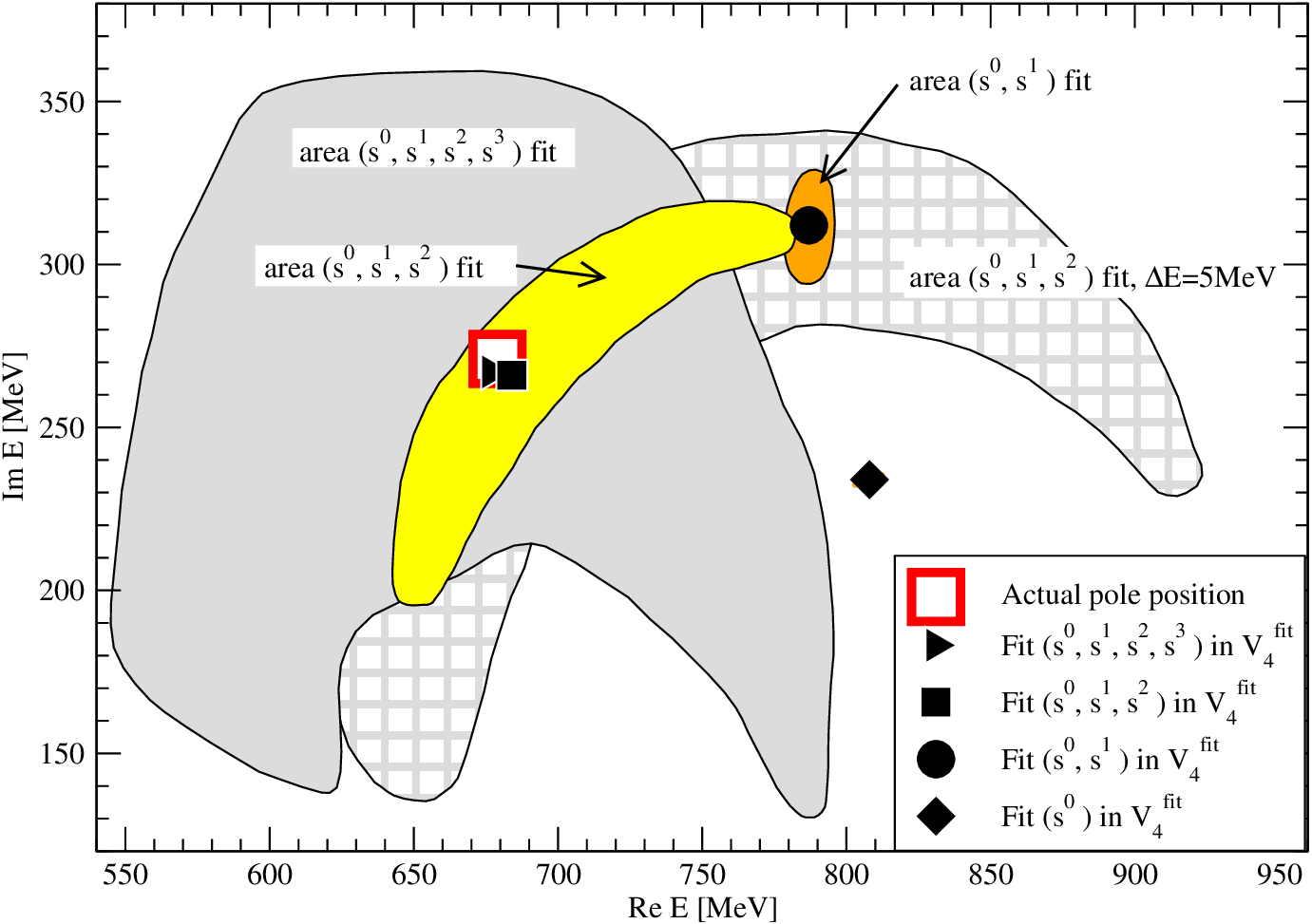}
\end{center}
\caption{Case of a low $\kappa(800)$ pole. Pole positions of the $\kappa$ (central values)
together with uncertainty areas, from fits to pseudo lattice-data. All symbols and areas are
defined as in figure~\ref{fig:pole_spread_high_pole}.
}
\label{fig:pole_spread_low_pole}
\end{figure}
Comparing with the previous case of a higher $\kappa$-pole, shown in
figure~\ref{fig:fit_high_pole}, the error bands both for levels and phase shifts are similar
for the respective ($s^0,\,s^1,\,s^2$) fits. 

In figure~\ref{fig:pole_spread_low_pole} the pole positions as extracted from the pseudo
lattice-data are shown for the different fit functions, in full analogy to the results shown
in figure~\ref{fig:pole_spread_high_pole}. The central values are indicated with the
symbols, the areas show the corresponding uncertainties.   
We can perceive the same behavior of the fits as in the previous case of a rather high
$\kappa$-pole position: expanding the fit potential to too low powers, wrong central values
are extracted in combination with too small uncertainty areas that give the impression of an
unrealistic precision. Once also the $s^2$-term is fitted, the central value is close to the
actual pole position. Including also the $s^3$-term in eq.~(\ref{vfit}) serves then to confirm
the result of the $(s^0,s^1,s^2)$ fit, because the central value is not changed any more and
only the uncertainty area grows. The realistic uncertainty is then again given by the yellow
shaded area from the $(s^0,s^1,s^2)$ fit. 


\subsection{Qualitative analysis using the lowest level}
\label{sec:lowest}
In the previous sections we have seen that relatively precise lattice data are needed to pin
down the pole position of the $\kappa(800)$. Still, one may wonder if with a few and low
precision data points from the lowest level, which is the one that can be most precisely
determined in actual lattice simulations, one can at least make qualitative statements about
the presence, absence, and maybe approximate position of the pole. 

\begin{figure}
\begin{center}
\includegraphics[width=0.8\textwidth]{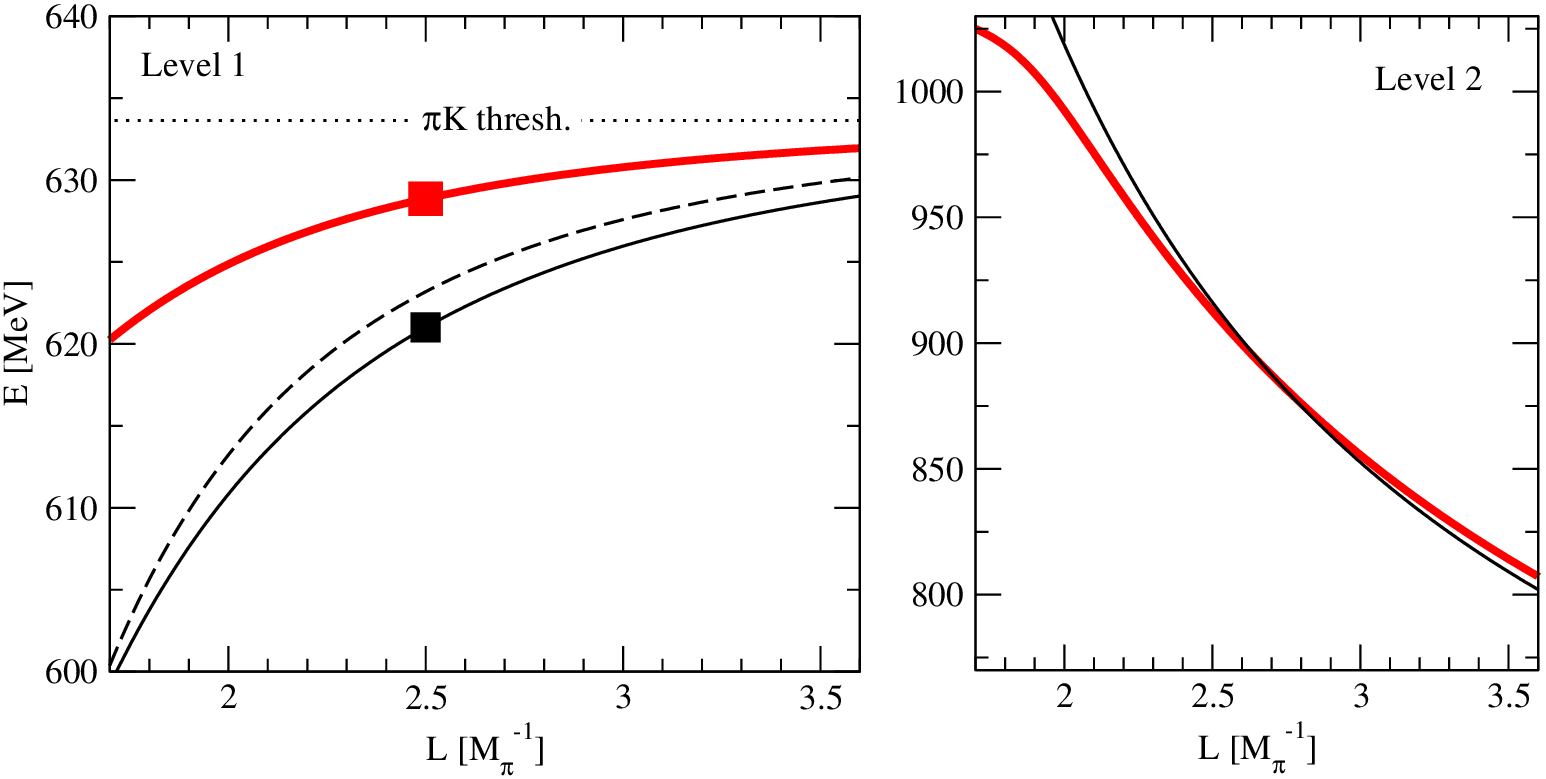}
\end{center}
\caption{Levels 1 and 2. Thick (red) solid lines: case of higher pole position from
section~\ref{sec:higher_pole}. Thin (black) solid lines: case of lower pole position from
section~\ref{sec:lowkappa}. Dashed line: Approximation of eq.~(\ref{approximation}) for the
case of lower pole position. The squares to the left indicate the measurements taken to
calculate the approximations to phase shifts and pole positions shown in
figure~\ref{fig:thres}.
}
\label{fig:lowest}
\end{figure}
In figure~\ref{fig:lowest} the levels 1 and 2 are shown. The case of a higher $\kappa(800)$
pole position [cf. section~\ref{sec:higher_pole}] is indicated with the thick (red) solid
line, the case of lower pole position [cf. section~\ref{sec:lowkappa}] with the thin (black)
solid line. As the figure shows, the largest differences of the two solutions are visible
for level 1, and those differences are enhanced to more than 10~MeV for the smallest $L$. In
contrast, level 2 shows only differences of a few MeV below $E\sim 900$~MeV. In the
following, we will study to which extent the low-$L$ enhancement of level 1 can help to draw
qualitative conclusions on the $\kappa(800)$ pole.

We start with the derivation of an approximate expression for the $L$-dependence of the
first level. In the present formulation the function $\tilde G$ of eq.~(\ref{tildegs}) can
be expanded in a Laurent series around the pole at threshold, 
\be \tilde
G=\frac{1}{4m_1\,m_2\,L^3}\,\frac{1}{E-(m_1+m_2)}+{\cal O}[E-(m_1+m_2)]^0 \  
\label{gpole}
\ee 
where we can ignore contributions from $G$ and $G_S$ in eq.~(\ref{tildegs}) of order
${\cal O}[E~-~(m_1+~m_2)]^{1/2}$. For an $S$-wave interaction, we can approximate the potential
$V$ close to threshold by a constant, $V^{\rm fit}=C_{\pi K}$. The levels are given by
$1-V^{\rm fit}\tilde G=0$ (the problem can be reduced to one channel for this exercise), so
that 
\be
E=m_1+m_2+\frac{C_{\pi K}}{4\,m_1\,m_2\,L^3} \ .
\label{approximation}
\ee
This is equivalent to the standard result for the scattering length, see
refs.~\cite{Luscher:1986pf,Luscher:1990ux,Beane:2003da}. There will always be a level close
to threshold with this functional behavior, because $\tilde G$ diverges at threshold and
always equals $(V^{\rm fit})^{-1}$ for some $E$ close to it. In particular, we read from
eq.~(\ref{approximation}) that if the interaction is attractive $(C_{\pi K}<0)$, the level
will be slightly below threshold, while for $C_{\pi K}>0$ it will be above. For the $S$-wave,
we observe exactly this behavior for all cases shown in figure~\ref{fig:levels}. Note that
once $C_{\pi K}$ is determined from lattice data using eq.~(\ref{approximation}), one can
relate it to the scattering length. With the isospin $I=1/2$ $S$-wave scattering length
defined as $\lim_{q_{\rm c.m.}\to 0} q_{\rm c.m.}\,\cot\delta_{1/2}^0=1/a_{1/2}^0$ and
$a_{1/2}^0=-{\rm Re}\,T_{1/2}^0(E=M_\pi+M_{K})/[8\pi\,(M_\pi+M_{K})]$ in the current
normalization of the amplitude $T$ we obtain 
\be
a_{1/2}^0=\frac{1}{8\pi\,E}\,\frac{C_{\pi K}}{C_{\pi K}\,G_{\pi K}-1}\bigg|_{E=M_\pi+M_{K}} 
\ .
\ee

Back to the original problem, the approximation of eq.~(\ref{approximation}) for level 1 is
shown in figure~\ref{fig:lowest} with the dashed line for the case of low $\kappa$ pole. The
agreement is quite good.  With $C_{\pi K}$ determined from eq.~(\ref{approximation}), we can
exploit the fit potential of eq.~(\ref{vfit}). The measured $C_{\pi K}$ fixes the parameter
$a$ of $V_4^{\rm fit}$, while $b,c,d$ are set to zero as they cannot be fixed, 
\be
V_4^{\rm fit}=V_2^{\rm thres.}\left(1-\frac{V_2^{\rm thres.}}{C_{\pi K}}\right)
={\rm const.},\quad  V_2^{\rm thres.}=V_2(s=(M_\pi+M_K)^2)\ .
\label{v4app}
\ee
Using eq.~(\ref{tv}), phase shifts and pole positions may be determined now. For the
numerical calculation, we have determined $C_{\pi K}$ by measuring the lowest level at $L=2.5
M_\pi^{-1}$, as shown by the squares in figure~\ref{fig:lowest}, and then using
eq.~(\ref{approximation}).
\begin{figure}
\begin{center}
\includegraphics[width=0.8\textwidth]{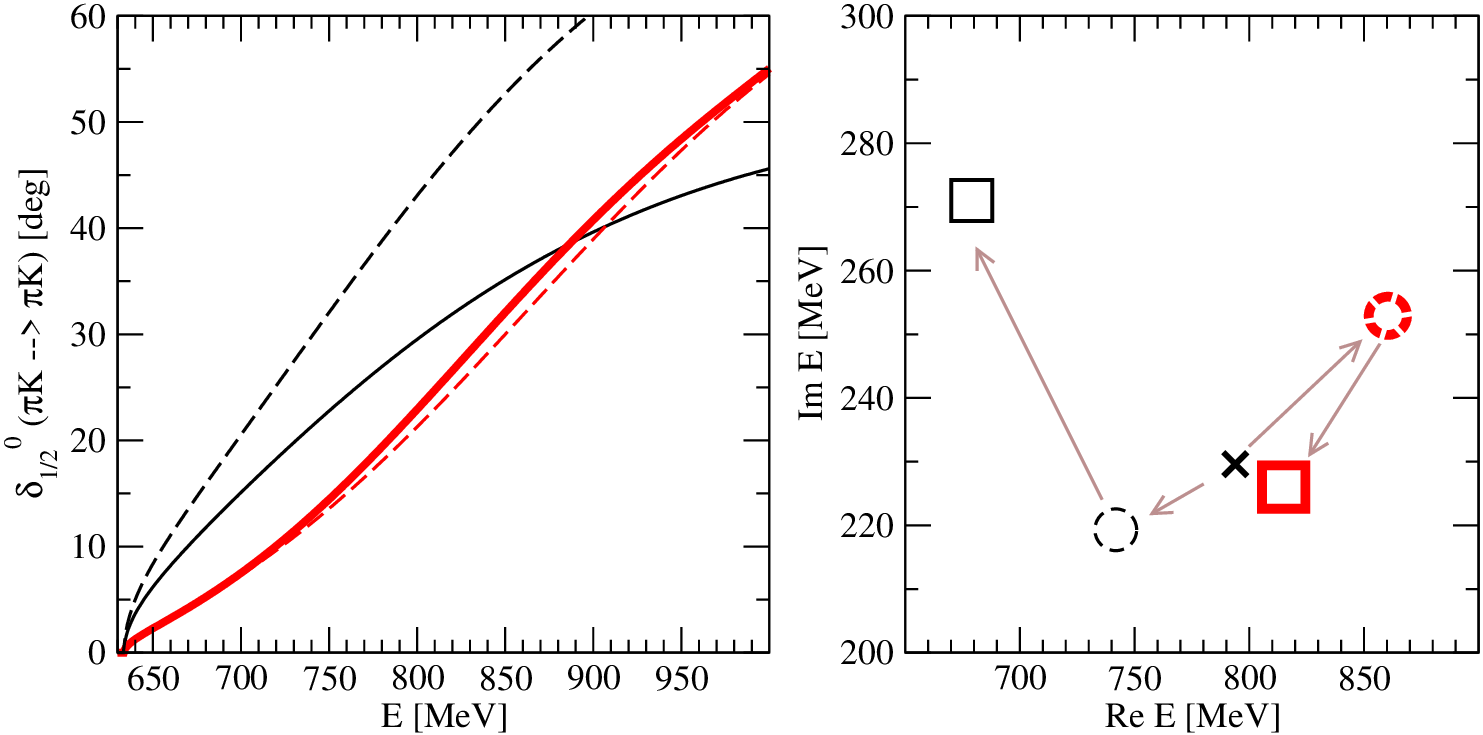}
\end{center}
\caption{Dashed lines/dashed circles: Approximations of phase shifts (left) and pole
positions (right) using only one data point from level 1 as shown by the squares in
figure~\ref{fig:lowest}. For comparison, the full results are shown: Thick solid (red) line
and thick (red) square: case of high $\kappa$ pole position. Thin solid (black) line and
thin (black) square: case of low $\kappa$ pole position.
}
\label{fig:thres}
\end{figure}
The result is shown in figure~\ref{fig:thres} for the two cases of higher and lower pole
position. In the left panel, the actual phase shifts (solid lines) and the ones determined
using eq.~(\ref{v4app}) are shown (dashed lines). Indeed, using only the information from the
lowest level, encoded in $C_{\pi K}$, the phase can be qualitatively described, in particular
for the case of a higher $\kappa$ pole. The pole positions are shown in the right panel of 
figure~\ref{fig:thres}. The actual pole positions are indicated with the squares, the ones
determined using eq.~(\ref{v4app}) are shown with the circles, and the pole position
obtained when setting $V_4^{\rm fit}=0$, i.e. the pole position from the LO calculation (in
one channel), is marked with the cross. Here, the situation is not as clear as for the
phases. In particular, the approximations of the actual pole positions, given by the method
proposed here (circles), show no systematic improvement compared to the case $V_4^{\rm
fit}=0$ (cross). In summary, measuring $V_4^{\rm fit}$ at one $L$ using the lowest
level improves the description of the phase close to threshold. However, beyond the
threshold region, clearly the higher order terms $\sim s,\,s^2$ in $V_4^{\rm fit}$ need to
be known to allow for a quantitative determination of the phase and the $\kappa$ pole.

To conclude this section, we come back to a question left open in the discussion of
figure~\ref{fig:levels}: why is there no level close to the lower thresholds of the $P$-wave
reactions, while there is always one for the $S$-wave reactions? For $P$-waves, the
potential close to threshold will be of the form $V=a'\,q_{\rm c.m.}^2$ with a nearly
constant $a'$.
Expanding $q_{\rm c.m.}$ as a function of $E$ one finds that the pole of $\tilde G$ (see eq.~(\ref{gpole})) is
precisely canceled in $1-V\tilde G=0$,
\be
1-V\tilde G\sim 1-\frac{a'}{2\,L^3\,(m_1+m_2)}
\ee
and there is no need for this expression to become zero, i.e. no need for the occurrence of
a level close to a $P$-wave threshold. The absence of avoided level crossing at the
respective upper thresholds, for the $P$-wave channels, can be explained similarly. 


\section{Extraction of other excited meson states}
\label{sec:other}
To put the previous results on the $\kappa(800)$ into context, in particular the required
high precision of the lattice data, in the following the $\sigma(600)$ is analyzed that has
similar properties. Also, the results for these $S$-wave resonances are contrasted with the
$P$-wave $K^*(892)$ and $\rho(770)$ states, as well as the repulsive $\pi K$ and $\pi\pi$
phase shifts.


\subsection{The \texorpdfstring{$\boldsymbol{\sigma(600)}$}{sigma(600)} 
at next-to-leading order}
\label{sec:sigma}

\begin{figure}
\begin{center}
\includegraphics[width=0.98\textwidth]{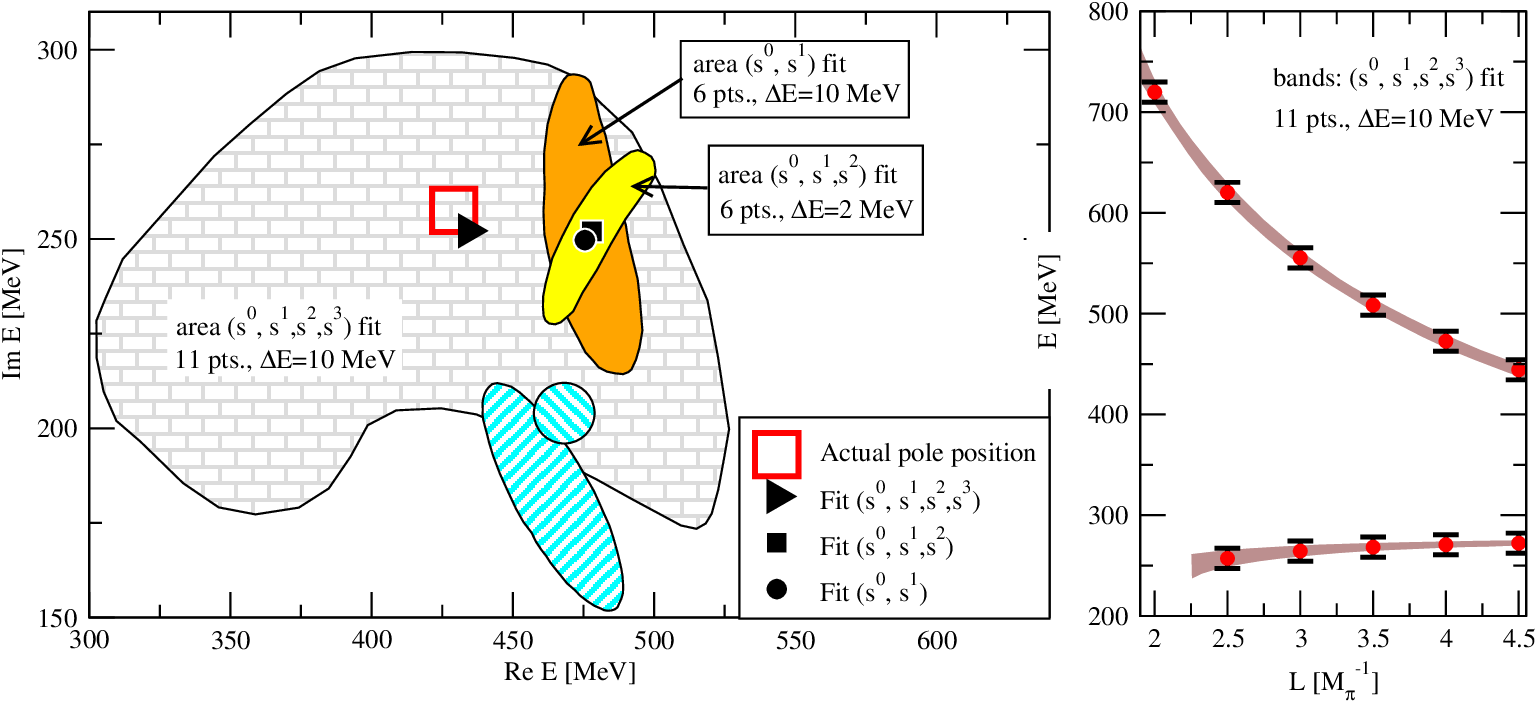}
\end{center}
\caption{Left: Pole reconstruction of the $\sigma(600)$. The (black) symbols show different
fits, the actual pole position is shown by the (red) open square. Uncertainty areas are
labeled. For comparison, the result from ref.~\cite{Doring:2011vk} is shown with the hatched
circle (pole position) and hatched uncertainty area, obtained from a $(s^0,\,s^1)$ fit.
Right: 6 data points from the upper level (level 2) plus 5 points from level 1, as used for
the fits. The data errors and uncertainty band for one fit is shown as indicated in the
figure.
}
\label{fig:sigmapole}
\end{figure}

In the previous sections, lattice data with an error of $2$~MeV were required to obtain a
sensible error on the pole position of the $\kappa(800)$, as shown with the (yellow shaded)
$(s^0,s^1,s^2)$-areas in figures~\ref{fig:pole_spread_high_pole} and
\ref{fig:pole_spread_low_pole}. This result is in certain contrast with the pole
reconstruction of the $\sigma(600)$ in ref.~\cite{Doring:2011vk} where a few data points
with a $10$~MeV error were found enough to obtain a reasonable determination of the pole. We
repeat the current analysis to put these differences into context.  For this, we use the
same values of $L$ as in ref.~\cite{Doring:2011vk} to generate 6 data points with a 10 MeV
error~\footnote{To generate the data, the global fit from section~\ref{sec:inifinitev} needed
to be slightly modified, because that solution contains, in the $(I,L,S)=(0,0,0)$ quantum
numbers, a pole in the $T$-matrix below the $\pi\pi$ threshold at $E=200$~MeV.  Note that
also the solution of ref.~\cite{Oller:1998hw} has a pole below threshold for these quantum
numbers. This is not noticeable in the phase shift, but the first lattice level, below the
$\pi\pi$ threshold, is affected at small $L$. Such  effects close to threshold are, apart
from being unphysical, difficult to accommodate in the expansion of the fit potential.}.
These data are fitted in a $(s^0,\,s^1)$ fit, like in ref.~\cite{Doring:2011vk}, with the
result for the pole position and uncertainty (orange area) shown in figure~\ref{fig:sigmapole}
to the left.
The corresponding result from ref.~\cite{Doring:2011vk} is shown with the hatched areas in
figure~\ref{fig:sigmapole}. As the figure shows, for the present result and the one from
ref.~\cite{Doring:2011vk}, the uncertainty is of quite similar size. However, the central
value in the present case (filled circle) is quite far from the actual position: Like in the
case of the $\kappa(800)$, there are now sizeable NLO terms $\sim~s^2$ that cannot be
accommodated in a simple  $(s^0,\,s^1)$ fit. Thus, to better reconstruct the original pole
position of the $\sigma(600)$, we have performed an $(s^0,\,s^1,\,s^2)$ fit, but with a
$2$~MeV error for the data (yellow area). As the figure shows, the corresponding uncertainty
area (yellow) is of similar size as for the case of the $\kappa(800)$ from
Figs.~\ref{fig:pole_spread_high_pole} and \ref{fig:pole_spread_low_pole}. Concluding, in
ref.~\cite{Doring:2011vk} a 10~MeV error on the data produces an uncertainty of similar size
as $2$~MeV data errors here, because the fit potential here is expanded to higher powers. As
figure~\ref{fig:sigmapole} shows, even the $(s^0,\,s^1,\,s^2)$ fit shows no agreement
of the central value with the actual pole position. The $(s^0,\,s^1,\,s^2, s^3)$ fit,
however, does. To minimize the uncertainty area for this case, in addition to the 6 points
from level 2, 5 points from level 1 are included in the fit, that would be known in an
actual lattice calculation anyway. As the figure shows, one can then provide again a 10~MeV
error to the data to obtain a large, but still limited uncertainty area~\footnote{Note that
according to the fit strategy formulated in section~\ref{sec:higher_pole}, in an analysis of
lattice data where the actual pole position is unknown, one would still have to carry out
the $(s^0,\,s^1,\,s^2, s^3, s^4)$ fit to confirm the pole position found in the
$(s^0,\,s^1,\,s^2, s^3)$ fit and accept the corresponding uncertainty area as the final
result. As we know the actual pole position for the pseudo-data, there is no need to perform
this task here.}. 


\subsection{P-wave states: Extraction of the \texorpdfstring{$\mathbf{K^*(892)}$}{K*(892)} 
and the \texorpdfstring{$\boldsymbol{\rho(770)}$}{rho(770)}}
\label{sec:pwave}
While the broad $\kappa(800)$ and $\sigma(600)$ $S$-wave resonances, discussed in the
previous sections, require relatively precise lattice data, some $P$-wave resonances are
expected to show clearer signals on the lattice: For $P$-wave the lowest lattice level
---i.e. the most accessible one--- is located already in the resonance region, in contrast
to the $S$-wave that has a level at the lowest threshold (see figure~\ref{fig:levels} and
section~\ref{sec:lowest}). Also, resonances like the $K^*(892)$ and $\rho(770)$ are much
narrower than the analyzed $S$-wave states and the corresponding $\pi K$ and $\pi\pi$
channels show only very weak coupling to the respective $\eta K$ and $\bar KK$
channels~\cite{Oller:1998hw}, making the analysis easier than in case of, e.g., the
$f_0(980)$ or $a_0(980)$ resonances~\cite{Bernard:2010fp,Doring:2011vk,Doring:2011ip}.
Lattice calculations and resonance extraction at finite volume for the $\rho(770)$ have
already allowed for a determination of its mass and the $\rho\pi\pi$ coupling constant, at a
pion mass of around twice the physical one, see, e.g., ref.~\cite{Lang:2011mn}. The
relativistic Breit-Wigner form, assumed in ref.~\cite{Lang:2011mn}, or effective range
formulae assumed in refs.~\cite{Gockeler:2008kc,Feng:2010es}, are adequate for the
$\rho(770)$, but the method presented here is a more general model-independent extraction of
the resonance properties. 

The $S$-wave fit potential from eq.~(\ref{vfit}) can be generalized to $P$-wave with minor
changes. Noting that the threshold behavior is given by $V^{\rm fit}_{P-{\rm wave}}\sim
q_{\rm c.m.}^2$, we can simply multiply this factor to the $S$-wave potential to obtain
\be
V^{\rm fit}_{P-{\rm wave}}=\left(\frac{V_2-V_4^{\rm fit}}{V_2^2}\right)^{-1},
\quad V_4^{\rm fit}=q_{\rm c.m.}^2\left[a+b(s-s_0)+c(s-s_0)^2+\cdots\right] 
\label{vfitp}
\ee
and $V_2$ is given by the $P$-wave projected LO term. 
\begin{figure}
\begin{center}
\includegraphics[width=0.98\textwidth]{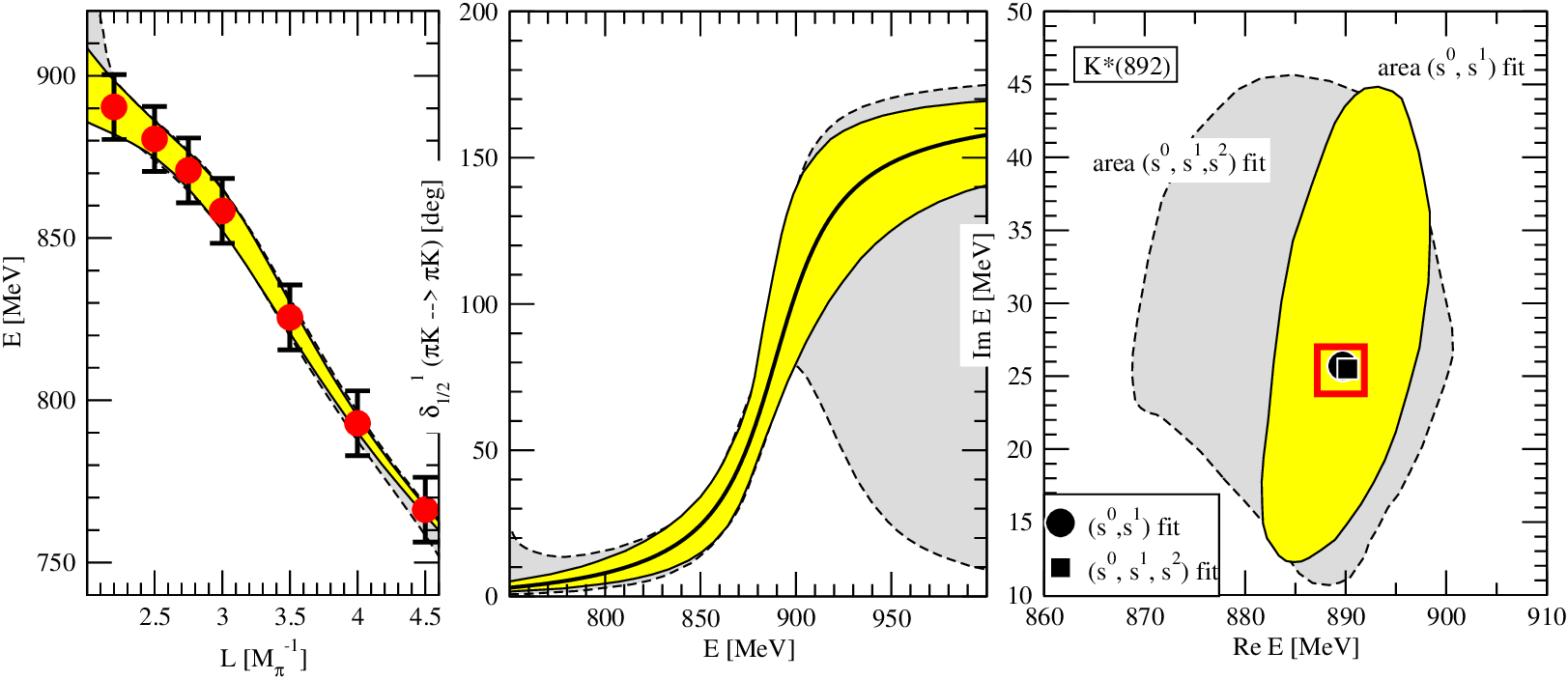}
\end{center}
\caption{Extraction of the $K^*(892)$ resonance using 7 data points from level 1 with
  $10$~MeV error (left). Center and right: Phase shifts and pole position, respectively. The
  (yellow) areas show the uncertainty of the $(s^0,s^1)$ fit, the (gray) areas the one of
  the $(s^0,s^1,s^2)$ fit. The actual phase [pole position] is indicated with the thick
  solid line [open (red) square]. The central values of the fitted pole position are
  indicated with the filled circle and the filled square.
}
\label{fig:Kstar}
\end{figure}

\begin{figure}
\begin{center}
\includegraphics[width=0.98\textwidth]{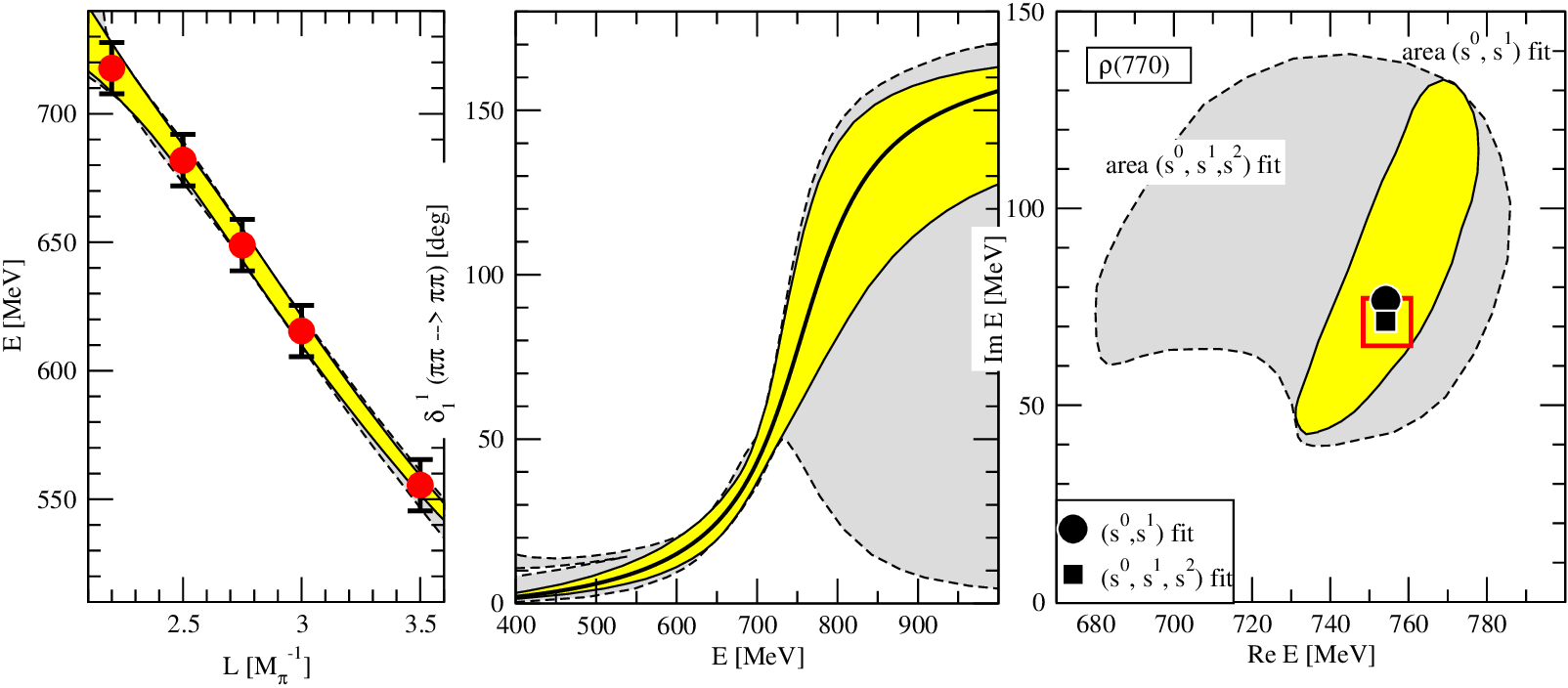}
\end{center}
\caption{Extraction of the $\rho(770)$ resonance using 5 data points from level 1 with
$10$~MeV error (left). Center and right: Phase shifts and pole position, respectively.
Labeling as in figure~\ref{fig:Kstar}.
}
\label{fig:rho}
\end{figure}
For the generation of pseudo-data in the $(\eta K,\,\pi K)$ $P$-wave channel, the global
solution determined in section~\ref{sec:inifinitev} has been used, while for the pseudo-data
of the $(\bar KK,\,\pi\pi)$  $P$-wave we use the solution of ref.~\cite{Oller:1998hw} that is
slightly better for this quantum number (dashed vs. solid line in
figure~\ref{fig:phase_shifts}). 

The analysis of the pseudo-data can be carried out as in the previous sections, with the
result for the $K^*(892)$ shown in figure~\ref{fig:Kstar} and for the $\rho(700)$ in
figure~\ref{fig:rho}. For both resonances, the $(s^0, s^1)$ fits lead to pole positions very
close to the actual ones, indicated with the open (red) squares. Note that contributions to
the $s^1$-term in $P$-wave correspond, among others, to terms of the form $ts$, $us$, $t^2$,
or $u^2$ as they appear in the NLO potential~\cite{Oller:1998hw}. The $(s^0, s^1, s^2)$ fits
return practically the same pole positions and much enlarged uncertainty areas. This
indicates that the $(s^0, s^1)$ fits and their uncertainty areas can be regarded as the
final results, according to the fit strategy formulated in section~\ref{sec:higher_pole}. 

The observed convergence for $(s^0, s^1)$ fits should be seen in comparison to the previous
results for the $S$-wave resonances: for the $\kappa(800)$, convergence occurs only for the
$(s^0, s^1,s^2)$ fit, while for the $\sigma(600)$ the potential needs to be expanded to even
one more power. Thus, the properties of the considered $P$-wave resonances are indeed much
easier to determine than their $S$-wave counterparts.

For the $K^*(892)$, seven data points up to values of $L=4.5\,M_\pi^{-1}$ have been fitted. 
As figure~\ref{fig:Kstar} shows, the (yellow) uncertainty area of
the levels is broad at low $L$ and narrow at large $L$. This indicates that the data at
small $L$ predominantly fix the fit result, while those at high $L$ may be dispensable.
Thus, for the analysis of the $\rho(770)$ the two data points at $L>3.5\,M_\pi^{-1}$ were
omitted as shown in figure~\ref{fig:rho}. Indeed, the $\rho(770)$ properties could be
determined to a similar quality as the ones of the $K^*(892)$, taking into account that the
larger width of the $\rho(770)$ automatically leads to an enlarged uncertainty area.

Thus, the analysis of pseudo-lattice data can help to identify regions of the lattice
spectrum that are particularly sensitive to resonance properties. Such findings can
contribute to minimize the required effort of actual lattice calculations. 


\subsection{The \texorpdfstring{$\boldsymbol{\pi K}$}{piK} and
\texorpdfstring{$\boldsymbol{\pi\pi}$}{pipi} repulsive phase shifts}
\begin{figure}
\begin{center}
\includegraphics[width=0.8\textwidth]{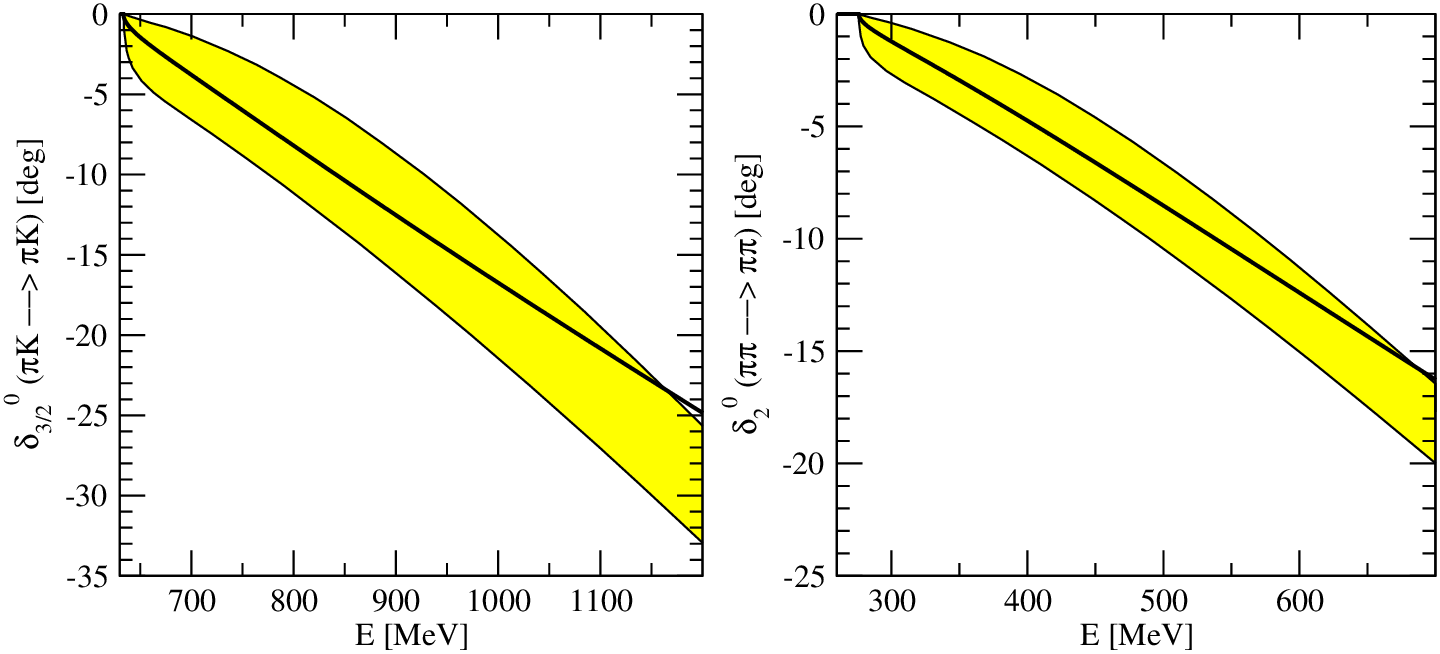}
\end{center}
\caption{Extraction of the repulsive $\pi K$ (left) and $\pi\pi$ phase shifts (right). The
thick solid lines show the actual phase shifts. The (yellow) bands show the extracted phase
shifts using one data point of the respective lowest level, at $L=2.5\,M_\pi^{-1}$ and with
a $10$~MeV error.
}
\label{fig:repulsive}
\end{figure}
To put the resonance analyses of the previous sections into context, we apply the same
method to phase shifts without resonances. The repulsive $\pi K$ isospin $3/2$ and $\pi\pi$
isospin $2$ phases show no structure and an almost linear energy dependence. As the $\pi\pi$
isospin $2$ phase shift provides a channel with maximal isospin, it is more accessible in
lattice calculations~\cite{Dudek:2010ew,Beane:2011sc} because disconnected diagrams are
absent (the same holds for the isospin 3/2 $\pi K$ S-wave). 
As figure~\ref{fig:levels} shows the second level is already at quite
high energies, for realistic values of $L$ and physical pion masses. Here, we restrict the
analysis to the lowest level, i.e. we proceed as in section~\ref{sec:lowest}. One could use
several data points from the lowest level, but, as shown in eq.~(\ref{approximation}), the
$L$-dependence is quite fixed and one data point at low values of $L$ is enough. For the
analysis, the data point is taken at $L=2.5\,M_\pi^{-1}$ with an assigned error of $10$~MeV,
using the values for the low energy constants of ref.~\cite{Oller:1998hw} that produce a
slightly better solution for the repulsive phases than the global one derived in this study
(dashed vs. solid lines in figure~\ref{fig:phase_shifts}).  As discussed in
section~\ref{sec:lowest}, using one data point allows only for a $(s^0)$ fit of the fit
potential of eq.~(\ref{vfit}). The results for both repulsive phase shifts are shown in
figure~\ref{fig:repulsive}. The actual phase shift lies inside the uncertainty area up to
energies of $500$~MeV above the $\pi K$ threshold and $400$~MeV above the $\pi\pi$ threshold,
respectively. Thus, due to the almost linear energy dependence of the phases, a single data
point is enough to reconstruct them up to quite high energies above the respective
thresholds. 


\subsection{Residues}
\label{sec:residues}
\begin{figure}
\begin{center}
\includegraphics[width=0.96\textwidth]{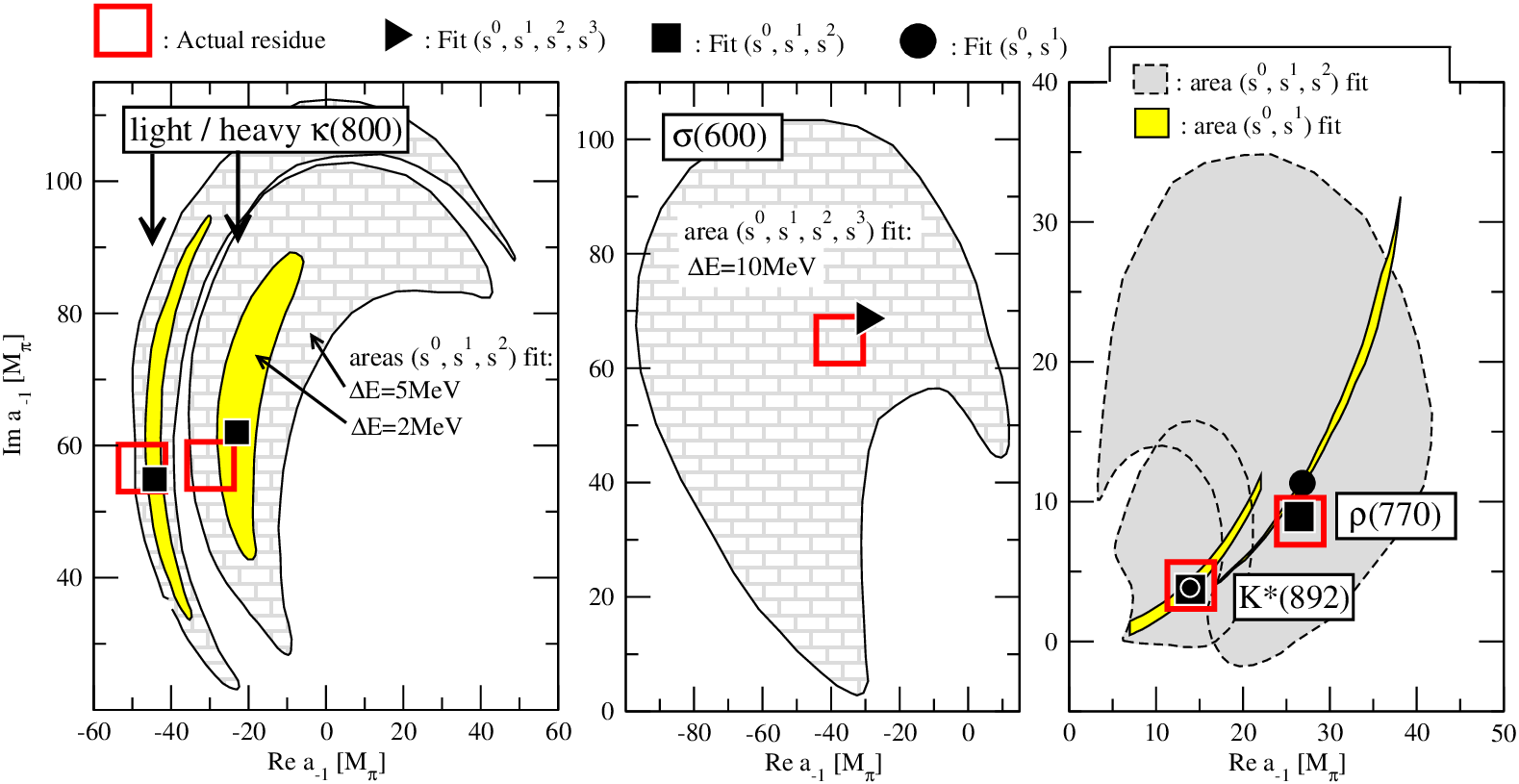}
\end{center}
\caption{Residues of the resonances. The labeling of symbols and areas is precisely as in
the figures of the respective resonances,
\ref{fig:pole_spread_high_pole},\ref{fig:pole_spread_low_pole}, \ref{fig:sigmapole},
\ref{fig:Kstar}, and \ref{fig:rho}. In the left panel, both cases of a heavy $\kappa(800)$
from section~\ref{sec:higher_pole} and a light $\kappa(800)$ from section~\ref{sec:lowkappa}
are shown. The central panel shows the residues of the $\sigma(600)$ discussed in
section~\ref{sec:sigma}, and in the right panel the residues of the $K^*(892)$ and
$\rho(770)$ from section~\ref{sec:pwave} are displayed.
}
\label{fig:residues}
\end{figure}
To finish the resonance analysis, the residues of all analyzed resonances with their
uncertainties are shown in figure~\ref{fig:residues}. The displayed residues are defined as 
\be T=\frac{a_{-1}}{E-z_0}+{\cal O}(E-z_0)^0 
\label{resi} 
\ee 
where $z_0$ is the pole position in the upper half plane and $T$ is the transition within the
lighter channel (on the second Riemann sheet), i.e. $\pi K\to~\pi K$ for the $\kappa(800)$ and
the $K^*(892)$; $\pi\pi\to\pi\pi$ for the $\sigma(600)$ and the $\rho(770)$. Note that the
residues for the poles at $z_0^*$ in the lower half plane are given as
$a_{-1}(z_0^*)=a_{-1}^*(z_0)$. For the values of the actual residues, shown by the open (red)
squares, the amplitude $T$ in eq.~(\ref{resi}) is given by eq.~(\ref{ia}), i.e. the solution
that was used to generate the pseudo-data. For the fitted residues (central values, shown by
filled (black) symbols, and areas), $T$ is given by eq.~(\ref{tv}), using eq.~(\ref{vfit}) or
(\ref{vfitp}) for the fit potential, depending on whether $S$- or $P$-wave states are analyzed.

The results shown in figure~\ref{fig:residues} suggest a similar interpretation as for the
respective pole positions shown in
figures~\ref{fig:pole_spread_high_pole},\ref{fig:pole_spread_low_pole}, \ref{fig:sigmapole},
\ref{fig:Kstar}, and \ref{fig:rho}: The residues of the broad $S$-wave resonances are much
more difficult to determine than the residues of the $P$-wave states and require higher
precision lattice data. For example, a $5$~MeV error for the lattice data of the
$\kappa(800)$ barely fixes the sign of the real part of the residue. In contrast, with only
a few lattice points from the lowest level and $10$~MeV error, the $(s^0,\,s^1)$ fits of the
$K^*(892)$ and $\rho(770)$ resonances lead to a very narrow error band on the respective
residues.


\section{Conclusions}
\label{sec:con}
For the reconstruction of the $\kappa(800)$ resonance from lattice data, we have first
performed a global fit to meson-meson partial wave data using the inverse amplitude method
based on next-to-leading order terms of the chiral expansion. This allows to determine the
low energy constants and to predict a pole position of $z_0=(815+i\,226)$~MeV for the
$\kappa(800)$. This solution was used to generate pseudo-lattice data that were subsequently
analyzed expanding a suitable fit potential systematically in powers of $s$. The key point
is that the lowest-order chiral interaction serves as explicit input to the fit potential,
considerably stabilizing the results. With a simple fit strategy the $\pi K$ phase and
$\kappa(800)$ pole position together with the respective uncertainties could be determined
in a model-independent way. To cover the possible pole positions of the $\kappa(800)$, we
have also analyzed the case of a lighter $\kappa(800)$. As has been shown, a comprehensive
resonance analysis requires rather high-precision lattice data. Still, using only one data
point from the lowest level, one can at least make qualitative statements about the
$\kappa(800)$. To put these results into context, we have also analyzed other channels using
the same technique, showing that the $P$-wave resonances $K^*(892)$ and $\rho(770)$ are much
easier to extract than the $S$-wave $\kappa(800)$ and $\sigma(600)$ resonances. The proposed
analysis method allows not only to quantify the required precision of lattice data, but also
helps to identify regions of the lattice spectrum that are particularly sensitive to
resonance properties, which can contribute to minimize the practical effort in actual
lattice simulations.


\acknowledgments
We thank J.~A.~Oller and J.~R.~Pel\'aez for providing details on their work and
B.~Kubis, E.~Oset, and A.~Rusetsky
for discussions. Financial support from the EU Integrated Infrastructure Initiative
HadronPhysics2 (contract number 227431) and DFG (SFB/TR 16, ``Subnuclear Structure of
Matter'') is gratefully acknowledged.


\end{document}